%% file: Main.tex
\documentclass{jfm}
\usepackage{xcolor}
\usepackage{ulem}
\usepackage{graphicx}
\usepackage{epstopdf, epsfig}
\usepackage{caption}
\usepackage{subcaption}
\usepackage{mathtools}
\def\req#1{(\ref{#1})} 

\shorttitle{AFC by RL with Partial Measurements}
\shortauthor{C. Xia, J. Zhang, E. C. Kerrigan and  G. Rigas}

\title{Active Flow Control for Bluff Body Drag Reduction Using Reinforcement Learning with Partial Measurements}

\author{Chengwei Xia\aff{1},
  Junjie Zhang\aff{1},
  Eric C. Kerrigan\aff{1,2},
  \and Georgios Rigas\aff{1}\corresp{\email{g.rigas@imperial.ac.uk}}}

\affiliation{\aff{1}Department of Aeronautics, Imperial College London,
London, SW7 2AZ, UK
\aff{2}Department of  Electrical and Electronic Engineering, Imperial College London, London, SW7 2AZ, UK}

\begin{document}

\maketitle

\begin{abstract}
Active flow control for drag reduction with reinforcement learning (RL) is performed in the wake of a 2D square bluff body at laminar regimes with vortex shedding. Controllers parameterised by neural networks are trained to drive two blowing and suction jets that manipulate the unsteady flow. RL with full observability (sensors in the wake) successfully discovers a control policy which reduces the drag by suppressing the vortex shedding in the wake. However, a non-negligible performance degradation ($\sim$ 50\% less drag reduction) is observed when the controller is trained with partial measurements (sensors on the body). To mitigate this effect, we propose an energy-efficient, dynamic, maximum entropy RL control scheme. First, an energy-efficiency-based reward function is proposed to optimise the energy consumption of the controller while maximising drag reduction. Second, the controller is trained with an augmented state consisting of both current and past measurements and actions, which can be formulated as a nonlinear autoregressive exogenous model, to alleviate the partial observability problem. Third, maximum entropy RL algorithms  (Soft Actor Critic and Truncated Quantile Critics) which promote exploration and exploitation in a sample efficient way are used and discover near-optimal policies in the challenging case of partial measurements. Stabilisation of the vortex shedding is achieved in the near wake using only surface pressure measurements on the rear of the body, resulting in similar drag reduction as in the case with wake sensors. The proposed approach opens new avenues for dynamic flow control using partial measurements for realistic configurations.

\end{abstract}


\input{Introduction.tex}

\input{Methodology.tex}

\input{Results.tex}

\input{Conclusion.tex}

\input{Appendix.tex}

\bibliographystyle{jfm}
\bibliography{jfm-instructions}

\end{document}

%% file: Introduction.tex
\section{Introduction}\label{sec:Intro}

Up to $50\%$ of total road vehicle energy consumption is due to aerodynamic drag  \citep{sudin_review_2014}.  
In order to improve vehicle aerodynamics, flow control approaches have been applied targeting the wake pressure drag, which is the dominant source of drag. Passive flow control has been applied \citep{choi2014aerodynamics} through geometry/surface modifications, e.g., boat tails \citep{lanser_aerodynamic_1991} and vortex generators \citep{lin_review_2002}. However, passive control designs do not adapt to environmental changes (disturbances, operating regimes), leading to sub-optimal performance under variable operating conditions. 
Active open-loop techniques, where pre-determined signals drive actuators, are typically energy inefficient since they target mean flow modifications. Actuators typically employed are synthetic jets \citep{glezer_synthetic_2002}, movable flaps \citep{beaudoin_bluff-body_2006,brackston_stochastic_2016} and plasma actuators \citep{corke_dielectric_2010}, among others. Since the flow behind vehicles is unsteady and subject to environmental disturbances and uncertainty, active feedback control is required to achieve optimal performance.  
However, two major challenges arise in feedback control design, which we aim to tackle in this study: (i) the flow dynamics are governed by the infinite-dimensional, nonlinear and non-local Navier-Stokes equations \citep{brunton2015closed}, and (ii) are partially observable in realistic applications due to sensor limitations. 
This study aims to tackle these challenges, particularly focusing on the potential of model-free control for a partially observable laminar flow, characterised by bluff body vortex shedding, as a preliminary step towards more complex flows and applications.

\subsection{Model-based active flow control} 

Model-based feedback control design requires a tractable model for the dynamics of the flow, usually obtained by data-driven or operator-driven techniques. 
Such methods have been applied successfully to control benchmark two-dimensional (2D) bluff body wakes, obtaining improved aerodynamic performance, e.g. vortex shedding suppression and drag reduction. 
For example, \cite{gerhard_model-based_2003} controlled the circular cylinder wake at low Reynolds numbers based on a low-dimensional model obtained from the Galerkin projection of Karhunen-Loeve modes on the governing Navier-Stokes equations. 
\cite{protas_linear_2004} applied Linear Quadratic Gaussian control to stabilise vortex shedding based on a F\"oppl point vortex model.
\cite{illingworth_model-based_2016} applied the Eigensystem Realization Algorithm as a system identification technique to obtain a reduced-order model of the flow and used robust control methods to obtain feedback control laws. 
\cite{jin_feedback_2020} employed resolvent analysis to obtain a low-order input-output model from the Navier-Stokes equations based on which feedback control was applied to suppress vortex shedding. 

Model-based flow control has also been applied at high Reynolds numbers to control dominant coherent structures (persisting spatio-temporal symmetry breaking modes) which contribute to drag, including unsteady vortex shedding   \citep{pastoor2008feedback, dahan_feedback_2012, dalla2017reducing, brackston_modelling_2018} and steady spatial symmetry breaking modes \citep{li2016feedback,   brackston_stochastic_2016}. 
For inhomogeneous flows in all three spatial dimensions, low-order models typically fail to capture the intractable and complex turbulent dynamics, leading inevitably to sub-optimal control performance when used in control synthesis.  

\subsection{Model-free active flow control by reinforcement learning} 

Model-free data-driven control methods bypass the above limitations by using input/output data from the dynamical system (environment) to learn the optimal control law (policy) directly without exploiting information from a mathematical model of the underlying process \citep{hou_data-driven_2009}.

Model-free reinforcement learning (RL) has been successfully used for controlling complex systems, for which obtaining accurate and tractable models can be challenging. RL learns a control policy based on observed states and generates control actions which maximise a reward by exploring and exploiting state-action pairs. The system dynamics governing the evolution of the states for a specific action (environment) are assumed to be a Markov Decision Process (MDP). The policy is parameterised by artificial neural networks as a universal function approximator that can be optimised to an arbitrary control function with any order of complexity.
RL with neural networks can also be interpreted as parameterised dynamic programming with the feature of universal function approximation \citep{bertsekas_reinforcement_2019}. Therefore, RL requires only input-output data from complex systems in order to discover control policies using model-free optimisation.

RL can effectively learn to control complex systems in various types of tasks, such as robotics \citep{kober_reinforcement_2013} and autonomous driving \citep{kiran_deep_2021}. In the context of chaotic dynamics related to fluid mechanics, \cite{bucci_control_2019} and \cite{zeng_symmetry_2021} applied RL to control the chaotic Kuramoto–Sivashinsky system. 
In the context of flow control for drag reduction, \citet{rabault_artificial_2019,rabault_accelerating_2019} used RL control for the first time in 2D bluff body simulations at a laminar regime. The RL algorithm discovered a policy that, using pressure sensors in the wake and near the body, drives blowing and suction actuators on the circular cylinder to decrease the mean drag and wake unsteadiness. 
\citet{tang2020robust} trained RL-controlled synthetic jets in the flow past a 2D cylinder at several Reynolds numbers, [100, 200, 300, 400], and achieved drag reduction in a range of Reynolds number from 60 to 400, showing the generalisation ability of RL active flow control.
\citet{paris_robust_2021} applied the ``S-PPO-CMA'' RL algorithm to control the wake behind a 2D cylinder and optimise the sensor locations in the near wake.
\citet{li_reinforcement-learning-based_2022} augmented and guided RL with global linear stability and sensitivity analyses in order to control the confined cylinder wake. They showed that if the sensors cover the wavemaker region, the RL is robust and successfully stabilises the vortex shedding. 
\citet{paris_reinforcement-learning-based_2023} proposed an RL methodology to optimise actuator placement in a laminar 2D flow around an airfoil, addressing the trade-off between performance and the number of actuators.
\citet{xu_reinforcement-learning-based_2023} used RL to suppress instabilities both in the Kuramoto-Sivashinsky system and 2D boundary layers, showing the effectiveness and robustness of RL control.
\citet{pino_comparative_2023} compared RL and genetic programming algorithms to global optimisation techniques for various cases, including the viscous Burger's equation and vortex shedding behind a 2D cylinder.
\citet{chen2023deep} applied RL in the flow control of vortex-induced vibration of a 2D square bluff body with various actuator layouts. The vibration and drag of the body were both reduced and mitigated effectively by RL policies.

Recently, RL has been used to control complex fluid systems, such as flows in turbulent regimes, in both simulations and experiments, addressing the potential of RL flow control in realistic applications.
\citet{fan_reinforcement_2020} extended RL flow control to a turbulent regime in experiments at Reynolds number of $O\left(10^5\right)$, achieving effective drag reduction by controlling the rotation speed of two cylinders downstream of a bluff body. RL successfully discovered the global optimal open-loop control strategy that was previously found from a laborious non-automated, systematic grid search. The experimental results were further verified by high-fidelity numerical simulations.
\citet{ren_applying_2021} examined RL-controlled synthetic jets in a weakly turbulent regime, demonstrating effective control at Reynolds number of 1000. This flow control problem of drag reduction of a 2D cylinder flow using synthetic jets was extended to Reynolds number of 2000 by \citet{varela2022deep}. In their work, RL discovered a strategy of separation delay via high-frequency perturbations to achieve drag reduction.
\citet{sonoda2023reinforcement} and \citet{guastoni2023deep} applied RL control in numerical simulations of turbulent channel flow and showed that RL control can outperform opposition control in this complex flow control task. 

RL techniques have been also applied to various flow control problems with different geometries, such as flow past a 2D cylinder \citep{rabault_artificial_2019}, vortex-induced vibration of a 2D square bluff body \citep{chen2023deep}, and a 2D boundary layer \citep{xu_reinforcement-learning-based_2023}. However, model-free RL control techniques also have several drawbacks compared to model-based control. For example, it is usually challenging to tune the various RL hyperparameters. Also,  model-free RL typically requires large amounts of training data through interactions with the environment, which makes RL expensive and infeasible for certain applications.
\color{black}
Further information about RL and its applications in fluid mechanics can be found in the reviews of \cite{garnier_review_2021} and \cite{vignon2023recent}.

\subsection{Maximum entropy reinforcement learning}

In RL algorithms, two major branches have been developed: ``on-policy'' learning and ``off-policy'' learning. RL algorithms can also be classified into value-based, policy-based, and actor-critic methods \citep{sutton_reinforcement_2018}. The actor-critic architecture combines advantages from both value-based and policy-based methods, so the state-of-the-art algorithms mainly use actor-critic architecture. 

The state-of-the-art on-policy algorithms include Trust Region Policy Optimization (TRPO, \cite{schulman_trust_2015}), Asynchronous Advantage Actor-Critic (A3C, \cite{mnih_asynchronous_2016}), and Proximal Policy Optimization (PPO, \cite{schulman_proximal_2017}). On-policy algorithms require fewer computational resources than off-policy algorithms, but they are demanding in terms of available data (interactions with the environment). They use the same policy to obtain experience in the environment and update with policy gradient, which introduces a high self-relevant experience that may restrict convergence to a local minimum and limit exploration. As the amount of data needed for training grows with the complexity of applications, on-policy algorithms usually require a long training time for collecting data and converging. 

By contrast, off-policy algorithms usually have both behaviour and target policies to facilitate exploration while retaining exploitation. The behaviour policy usually employs stochastic behaviour to interact with an environment and collect experience, which is used to update the target policy. There are many off-policy algorithms emerging in the past decade, such as Deterministic Policy Gradient (DPG, \cite{silver_deterministic_2014}), Deep Deterministic Policy Gradient (DDPG, \cite{lillicrap_continuous_2015}), Actor-Critic with Experience Replay (ACER, \cite{wang_sample_2016}), Twin Delayed Deep Deterministic Policy Gradient (TD3, \cite{fujimoto_addressing_2018}), Soft Actor-Critic (SAC, \cite{haarnoja_soft_2018-1,haarnoja_soft_2018}) and Truncated Quantile Critics (TQC, \cite{kuznetsov_controlling_2020}). 
Due to the behaviour-target framework, off-policy algorithms are able to exploit past information from a replay buffer to further increase sample efficiency. This ``experience replay'' suits a value-function-based method \citep{mnih_human-level_2015}, instead of calculating the policy gradient directly. Therefore, most of the off-policy algorithms implement an actor-critic architecture, e.g. SAC. 

One of the challenges of off-policy algorithms is the brittleness in terms of convergence. \cite{sutton_convergent_2008,sutton_fast_2009} tackled the instability issue of off-policy learning with linear approximations. They used a Bellman-error-based cost function together with stochastic gradient descent (SGD) to ensure the convergence of learning. \cite{maei_convergent_2009} further extended this method to nonlinear function approximation using a modified temporal difference algorithm. However, some algorithms nowadays still experience the problem of brittleness when using improper hyperparameters. Adapting these algorithms for control in various environments is sometimes challenging, as the learning stability is sensitive to their hyperparameters, such as DDPG \citep{duan_benchmarking_2016,henderson_deep_2018}.

To increase sample efficiency and learning stability, off-policy algorithms were developed within a maximum entropy framework \citep{ziebart_maximum_2008,haarnoja_reinforcement_2017}, known as ``maximum entropy reinforcement learning''. Maximum entropy RL solves an optimisation problem by maximizing the cumulative reward augmented with an entropy term. In this context, the concept of entropy was first introduced by \citet{shannon1948mathematical} in the information theory. The entropy quantifies the uncertainty of a data source, which is extended to the uncertainty of the outputs of stochastic neural networks in the RL framework. During the training phase, the maximum entropy RL maximises rewards and entropy simultaneously to improve control robustness \citep{ziebart2010modeling} and increase exploration via diverse behaviours \citep{haarnoja_reinforcement_2017}. Further details about maximum entropy RL and two particular algorithms used in the present work (SAC and TQC) are introduced in \S\ref{subsec:SACTQC}.

\subsection{Partial measurements and POMDP}

In most RL flow control applications, RL controllers have been assumed to have {full-state} information {(the term ``state'' is in the context of control theory)} or a sensor layout without any limitations on the sensor locations. In this study, it is denoted as ``full measurement'' (FM) when measurements contain full-state information. In practical applications, measurements are typically obtained on the surface of the body (e.g. pressure taps), and only partial-state information is available {due to the missing downstream evolution of the system dynamics}. This is denoted as ``partial measurement'' (PM), comparatively. 
PM can lead to control performance degradation compared to FM because the sensors are restricted from observing enough information from the flowfield. 
In the control of vortex shedding, full stabilisation can be achieved by placing sensors within the wavemaker region of bluff bodies, which is located approximately at the end of the recirculation region. In this case, full-state information regarding the vortex shedding is available to sensors. Placing sensors far from the recirculation region, for example, on the rear surface of the bluff body (denoted as PM in this work), introduces a convection delay of vortex shedding sensing and partial observation of the state of the system.
 
In the language of RL, control with PM can be described as a Partially Observable Markov Decision Process (POMDP)\citep{cassandra_survey_1998} instead of an MDP. 
In POMDP problems, the best stationary policy can be arbitrarily worse than the optimal policy in the underlying MDP \citep{singh1994learning}. In order to improve the performance of RL with POMDP, additional steps are required to reduce the POMDP problem to an MDP problem. This can be done trivially by using an augmented state, known as ``sufficient statistic'' \citep{bertsekas_dynamic_2012}, i.e. augmenting the state vector with past measurements and actions \citep{bucci_control_2019,wang2023dynamic}, or  Recurrent Neural Networks (RNN), such as Long-Short Term Memory (LSTM) \citep{verma2018efficient}. Theoretically, LSTM networks and augmented state approaches can yield comparable performance in partially observable problems \citep[see][Supplementary]{cobbe2020leveraging}. Practically, the augmented state methodology provides notable benefits, including reduced training complexity and ease in parameter tuning, provided that the control state dynamics are tractable and short-term correlated.

In the specific case for which flowfield information is available, POMDP can also be reduced to an MDP by flow reconstruction techniques based on supervised learning. For instance, \citet{bright2013compressive} estimates the full state based on a library containing the reduced order information from the full flowfield. However, there might be difficulties in constructing such a library as the entire flowfield might not be available in practical applications.

\subsection{Contribution of the present work}

The present work uses RL to discover control strategies of partially observable fluid flow environments {without access to the full flow-field/state measurements}. Fluid flow systems typically exhibit more complex sampling in higher dimensional observation space compared to other physical systems, necessitating a robust exploration strategy and rapid convergence in the optimisation process. To address these challenges, we employ off-policy-maximum entropy RL algorithms (SAC and TQC) that efficiently identify {nearly} optimal policies in the large action space inherent to fluid flow systems, especially for cases with partial measurements and observability.

We aim to achieve two objectives related to RL flow control for bluff body drag reduction problems. First, we aim to improve the RL control performance in a PM environment by reducing a POMDP problem to an MDP problem. More details about this method are introduced in 
\S\ref{subsec:PM_Dynamic}. Second, we present investigations on different reward functions and key hyperparameters to develop an approach that can be adapted to a broader range of flow control applications. We demonstrate the proposed framework and its capability to discover {nearly} optimal feedback control strategies in the benchmark laminar flow of a square 2D bluff body with fixed separation at the trailing edge, using sensors only on the {downstream surface} of the body. 

The article is structured as follows. In Section \S\ref{sec:Method}, the RL framework is presented, which consists of the SAC and TQC optimisation algorithms interacting with the flow simulation environment. A hyperparameter-free reward function is proposed to optimise the energy efficiency of the dynamically controlled system. Exploiting past action-state information converts the POMDP problem in a PM environment to an MDP, enabling the discovery of nearly optimal policies.
Results are presented and discussed in Section \S\ref{sec:Results}. The convergence study of RL is first introduced. The degradation of RL control performance in PM environments (POMDP) is presented, and the improvement is addressed by exploiting a sequence of past action-measurement information. At the end of this section, we compare the results from TQC with SAC, addressing the advantages of using TQC as an improved version of SAC.
In Section \S\ref{sec:Conclusions}, we provide conclusions for the current research and discuss future research directions. 

%% file: Methodology.tex
\section{Methodology}\label{sec:Method}

\begin{figure}
  \centerline{\includegraphics[width=13cm]{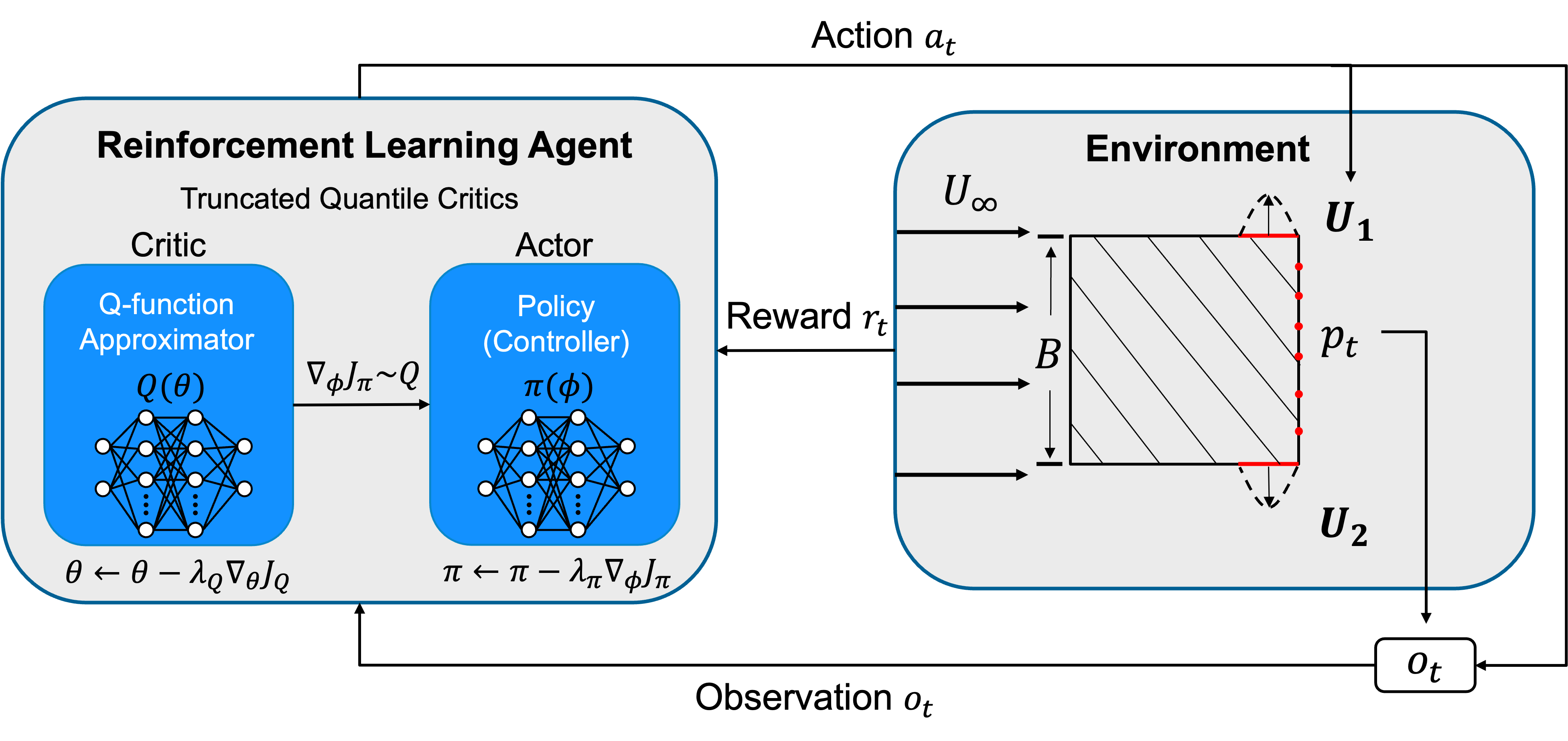}}
  \caption{Reinforcement learning framework. The RL agent, flow environment and the interaction between them are demonstrated. The partial measurement (PM) case is shown, where sensors are located on the {downstream surface} of the square bluff body. 64 sensors are placed by default, and the red dots only show a demonstration with a reduced number of sensors. Two jets located upstream of the rear separation points are trained to control the unsteady wake dynamics (vortex shedding).}
\label{fig:RL_framework}
\end{figure}

We demonstrate the RL drag reduction framework on the flow past a 2D square bluff body at laminar regimes characterised by two-dimensional vortex shedding.
We study the canonical flow behind a square bluff body due to the fixed separation of the boundary layer at the rear surface, which is relevant to road vehicle aerodynamics.
Control is applied by two jet actuators at the rear edge of the body before the fixed separation and partial- or full-state observations are obtained from pressure sensors on the {downstream surface} or near wake region, respectively.
The RL agent handles the optimisation, control and interaction with the flow simulation environment, as shown in figure \ref{fig:RL_framework}. The instantaneous signals $a_t$, $o_t$ and $r_t$ denote actions, observations and rewards at time step $t$.

Details of the flow environment are provided in \S\ref{subsec:Flow}. The SAC and TQC RL algorithms used in this work are introduced in \S\ref{subsec:SACTQC}. The reward functions based on optimal energy efficiency are presented in \S\ref{subsec:Reward}. The method to convert a POMDP to an MDP by designing a dynamic  {feedback} controller for achieving nearly optimal RL control performance is discussed in \S\ref{subsec:PM_Dynamic}.

\subsection{Flow environment}\label{subsec:Flow}
The environment is a 2D Direct Numerical Simulation (DNS) of the flow past a square bluff body of height $B$. The velocity profile at the inflow of the computational domain is uniform with freestream velocity $U_\infty$.  {Length quantities are non-dimensionalised with the bluff body height $B$ and velocity quantities are non-dimensionalised with the freestream velocity $U_\infty$. Consequently, time is non-dimensionalised with $B/U_\infty$.} The Reynolds number, defined as $Re = U_{\infty} B/\nu$, is $100$. The computational domain is rectangular with boundaries at  $(-20.5,26.5)$ in the streamwise $x$ direction and $(-12.5,12.5)$ in the transverse $y$ direction. The centre of the square bluff body is at $(x,y) = (0,0)$. The flow velocity is denoted as $\boldsymbol{u} = (u,v)$ where $u$ is the velocity component in the $x$ direction and $v$ in the $y$ direction.

The DNS flow environment is simulated using FEniCS and the Dolfin library \citep{logg_dolfin_2012}, based on the implementation of \cite{rabault_artificial_2019,rabault_accelerating_2019}. The incompressible unsteady Navier-Stokes equations are solved using a finite element method and the incremental pressure correction scheme \citep{goda_multistep_1979}. The DNS time step is $dt = 0.004$.  {More simulation details are presented in Appendix \ref{App:Sim_details}, including the mesh and boundary conditions.}

Two blowing and suction jet actuators are placed on the top and bottom surfaces of the bluff body before separation. The velocity profile $\boldsymbol{U_{j}}$ of the two jets ($j=1, 2$; 1 for the top jet and 2 for the bottom jet) is defined as
\begin{equation}
\boldsymbol{U_{j}} = \left(0, \quad \frac{3 Q_{j}}{2 w}\left[1-\left(\frac{2 x_j-L+w}{w}\right)^{2}\right]\right),
\label{eq:jet_u}
\end{equation}
where $Q_j$ is the mass flow rate of the jet $j$, and  {$L=B$ is the streamwise length of the body.  The width of the jet actuator is $w=0.1$, and the jets are located at $x_j \in [\frac{L}{2}-w,\frac{L}{2}]$, $y_j = \pm \frac{B}{2}$.}  
A zero mass flow rate condition of the two jets enforces momentum conservation as
\begin{equation}
Q_{1}+Q_{2}=0.
\label{eq:zero_mass}
\end{equation}
The mass flow rate of the jets is also constrained as $|Q_j|\leqslant0.1$ to avoid excessive actuation.

In PM environments, $N$ vertically equispaced pressure sensors are placed on the  {downstream surface} of the bluff body, the coordinates of which are given by
\begin{equation}
\boldsymbol{P_{surf,k}}=\left(\frac{B}{2},\frac{-B}{2}+ k \frac{B}{N+1}\right),
\label{eq:Probe_base}
\end{equation}
where $k = 1,2....,N$,  {and $N = 64$ unless specified.}
In FM environments, $64$ pressure sensors are placed in the wake region with a refined bias close to the body. The locations of sensors in the wake are defined with sets $\boldsymbol{x_s} = \left[0.25, 0.5, 1.0, 1.5, 2.0, 3.0, 4.0, 5.0\right]$ and $\boldsymbol{y_s} = \left[-1.5, -1.0, -0.5, -0.25, 0.25, 0.5, 1.0, 1.5\right]$, following the formula
\begin{equation}
\boldsymbol{P_{wake,i,j}}=\left(\frac{B}{2} + x_{s,i}, y_{s,j}\right),
\label{eq:Probe_wake}
\end{equation}
where $i = 1,2....,8$ and $j = 1,2....,8$.

The bluff body drag coefficient $C_{D}$ is defined as
\begin{equation}
C_{D}=\frac{F_{D}}{\frac{1}{2} \rho_{\infty} {U_{\infty}}^{2} B},
\label{eq:CD}
\end{equation}
and the lift coefficient $C_{L}$ as
\begin{equation}
C_{L}=\frac{F_{L}}{\frac{1}{2} \rho_{\infty} {U_{\infty}}^{2} B},
\label{eq:CL}
\end{equation}
where $F_{D}$ and $F_{L}$ are the drag and lift forces, defined as the surface integral of the pressure and viscous forces on the bluff body with respect to the $x$ and $y$ coordinates, respectively.

\subsection{Maximum entropy reinforcement learning of MDPs} \label{subsec:SACTQC}

RL can be defined as policy search in a Markov Decision Process (MDP), with a tuple $(\mathcal{S}, \mathcal{A}, \mathcal{P}, \mathcal{R})$ where $\mathcal{S}$ is a set of states, and $\mathcal{A}$ is a set of actions. $\mathcal{P}\left(s_{t+1} \mid s_t, a_t\right)$ is a state transition function that contains the probability from current state $s_t$ and action $a_t$ to the next state $s_{t+1}$. $\mathcal{R}(s, a)$ is a reward function (cost function) to be maximised. The RL agent collects data as states $s_t \in \mathcal{S}$ from the environment, and a policy $\pi\left(a_t \mid s_t\right)$ executes actions $a_t \in \mathcal{A}$ to drive the environment to the next state $s_{t+1}$. 

A state is considered to have the Markov property if the state at time $t$ retains all the necessary information to determine the future dynamics at $t+1$, without any information from the past \citep{sutton_reinforcement_2018}. This property can be presented as
\begin{equation}
\mathcal{P}\left\{r_{t+1}, s_{t+1} \mid s_{t}, a_{t}\right\} \equiv \mathcal{P}\left\{r_{t+1}, s_{t+1} \mid s_{0}, a_{0}, r_{1}, \ldots, s_{t-1}, a_{t-1}, r_{t}, s_{t}, a_{t}\right\}.
\label{eq:markov_property}
\end{equation}
In the present flow control application, the control task can be regarded as an MDP if observations $o_t$ contain full-state information, i.e. $o_t = s_t$, and satisfy  \req{eq:markov_property}.

SAC and TQC are two maximum entropy RL algorithms used in the present work. TQC is used by default since it is regarded as an improved version of SAC.
The maximum entropy RL generally maximises
\begin{equation}
J\left(\pi\right) = \sum_{t=0}^T \mathbb{E} \left[r_t\left(s_t, a_t\right)+\alpha \mathcal{H}\left(\pi\left(\cdot \mid s_t\right)\right)\right],
\label{eq:RL_Optimize}
\end{equation}
where $r_t$ is the reward (reward functions given in \S\ref{subsec:Reward}), and $\alpha$ is an entropy coefficient (known as ``temperature''), which controls the stochasticity (exploration) of the policy. For $\alpha=0$, the standard maximum reward optimisation in conventional reinforcement learning is recovered. The probability distribution (Gaussian distribution by default) of a stochastic policy is denoted by $\pi\left(\cdot \mid s_t\right)$. The entropy of $\pi\left(\cdot \mid s_t\right)$ is by definition \citep{shannon1948mathematical}
\begin{equation}
\mathcal{H}\left(\pi\left(\cdot \mid s_t\right)\right) = \mathbb{E} \left[ -\log\pi\left(\cdot \mid s_t\right)\right] = -\int_{\hat{a}_t} \pi\left(\hat{a}_t \mid s_t\right) \log \pi\left(\hat{a}_t \mid s_t\right) d \hat{a}_t,
\label{eq:entropy}
\end{equation}
where the term $-\log\pi$ quantifies the uncertainty contained in the probability distribution, and $\hat{a}_t$ is a distribution variable of the action $a_t$. Therefore, by calculating the expectation of $-\log\pi$, the entropy increases when the policy has more uncertainties, i.e. the variance of $\pi\left(\hat{a}_t \mid s_t\right)$ increases.

SAC is developed based on Soft Policy Iteration (SPI) \citep{haarnoja_soft_2018}. SPI uses a soft Q-function to evaluate the value of a policy and optimises the policy based on its value. The soft Q-function is calculated by applying a Bellman backup operator $\mathcal{T}^\pi$ as
\begin{equation}
\mathcal{T}^\pi Q\left({s}_t, {a}_t\right) \triangleq r_t\left({s}_t, {a}_t\right)+\gamma \mathbb{E}_{\mathbf{s}_{t+1} \sim \mathcal{P}}\left[V\left({s}_{t+1}\right)\right],
\label{eq:Bellman}
\end{equation}
where $\gamma$ is a discount factor (here $\gamma=0.99$), and $V\left({s}_{t+1}\right)$ satisfies
\begin{equation}
V\left({s}_t\right)=\mathbb{E}_{{a}_t \sim \pi}\left[Q\left({s}_t, {a}_t\right)-\log \pi\left({a}_t \mid {s}_t\right)\right].
\end{equation}
The target soft Q-function can be obtained by repeating $Q = \mathcal{T}^\pi Q$, and the proof of convergence can be referred to as Soft Policy Evaluation (Lemma 1) in \cite{haarnoja_soft_2018}. With soft Q-function rendering values for the policy, the policy optimisation is given as Soft Policy Improvement (Lemma 2 in \cite{haarnoja_soft_2018}). 

In SAC, a stochastic soft Q-function $Q_\theta\left({s}_t, {a}_t\right)$ and a policy $\pi_\phi\left({a}_t \mid {s}_t\right)$ are parameterised by artificial neural networks $\theta$ (critic) and $\phi$ (actor), respectively. 
During training, $Q_\theta\left({s}_t, {a}_t\right)$ and $\pi_\phi\left({a}_t \mid {s}_t\right)$ are optimised with stochastic gradients ${\nabla}_{\theta}J_Q(\theta)$ and $\nabla_\phi J_\pi(\phi)$ designed corresponding to Soft Policy Evaluation and Soft Policy Improvement respectively (see equation (6) and (10) in \cite{haarnoja_soft_2018}). With these gradients, SAC updates the critic and actor networks by
\begin{equation}
\theta \leftarrow \theta-\lambda_Q {\nabla}_{\theta} J_Q\left(\theta\right),
\label{eq:Q_update}
\end{equation}
\begin{equation}
\phi \leftarrow \phi-\lambda_\pi \nabla_\phi J_\pi(\phi),
\label{eq:Pi_update}
\end{equation}
where $\lambda_Q$ and $\lambda_\pi$ are the learning rates of Q-function and policy, respectively.
Typically, two Q-functions are trained independently, and then the minimum of the Q-functions is brought into the calculation of stochastic gradient and policy gradient. This method is also used in our work to increase the stability and speed of training.
SAC also supports automatic adjustment of temperature $\alpha$ by optimisation, 
\begin{equation}
\alpha^*=\arg \min _{\alpha} \mathbb{E}_{{a}_t \sim \pi^*}\left[-\alpha \log \pi^*\left({a}_t \mid {s}_t ; \alpha\right)-\alpha \overline{\mathcal{H}}\right].
\label{eq:temp_optimize}
\end{equation}
This adjustment transforms a hyperparameter-tuning challenge into a trivial optimisation problem  \citep{haarnoja_soft_2018}.

TQC \citep{kuznetsov_controlling_2020} can be regarded as an improved version of SAC as it alleviates the overestimation bias of the Q-function on the basic algorithm of SAC. 
TQC adapts the idea of distributional reinforcement learning with quantile regression, i.e. QR-DQN \citep{dabney_distributional_2018}, to format the return function $R(s, a):=\sum_{t=0}^{\infty} \gamma^t r_t\left(s_t, a_t\right)$ into a distributional representation with Dirac delta functions as 
\begin{equation}
R_{\psi}(s, a):=\frac{1}{M} \sum_{m=1}^M \delta\left(z_{\psi}^m(s, a)\right),
\label{eq:R_distribution}
\end{equation}
where $R(s, a)$ is parameterised by $\psi$, and $R_{\psi}(s, a)$ is converted into a summation of $M$ ``atoms'' as $z_{\psi}^m(s, a)$. Here only one approximation of $R(s, a)$ is used for demonstration.
Then, only $k$ smallest atoms of $z_{\psi}^m(s, a)$ are preserved as a truncation to obtain
truncated atoms 
\begin{equation}
y_i(s, a):=r(s, a)+\gamma\left[z_{\psi}^i\left(s^{\prime}, a^{\prime}\right)-\alpha \log \pi_\phi\left(a^{\prime} \mid s^{\prime}\right)\right], \quad i \in[1 . . k],
\label{eq:truncated_atoms}
\end{equation}
where $s^{\prime} \sim \mathcal{P}(\cdot \mid s, a)$ and $ a^{\prime} \sim \pi\left(\cdot \mid s^{\prime}\right)$. The truncated atoms form a target distribution as 
\begin{equation}
Y(s, a):=\frac{1}{k} \sum_{i=1}^{k} \delta\left(y_i(s, a)\right),
\label{eq:target_distribution}
\end{equation}
and the algorithm minimises the 1-Wasserstein distance between the original distribution $R_{\psi}(s, a)$ and the target distribution $Y(s, a)$ to obtain a truncated quantile critic.
Further details, such as the design of loss functions and the pseudocode of TQC, can be found in \cite{kuznetsov_controlling_2020}.

In this work,  {SAC and TQC are implemented based on Stable-Baselines3 and Stable-Baselines3-Contrib \citep{stable-baselines3}.} The RL interaction runs on a longer time step $t_a = 0.5$ compared to the numerical time step $dt$. This means RL-related data $o_t$, $a_t$ and $r_t$ are sampled every $t_a$ time interval. With a different numerical step and an RL step, control actuation $c_{n_s}$ for every numerical step should be distinguished from action $a_t$ in RL. There are $\frac{t_a}{dt}=125$ numerical steps between two RL steps, and control actuation is applied based on a first-order-hold function as
\begin{equation}
c_{n_s}= a_{t-1} + (a_t - a_{t-1})\frac{n_s dt}{t_a},
\label{eq:FOH_action}
\end{equation}
where $n_s$ denotes the number of numerical steps after generating the current action $a_t$ and before the next action $a_{t+1}$ generated. Equation \req{eq:FOH_action} smooths the control actuation with linear interpolation to avoid numerical instability.
Unless specified, the neural network configuration is set as 3 layers of 512 neurons for both actor and critic. The entropy coefficient in \req{eq:RL_Optimize} is initialised to $0.01$ and automatically tuned based on \req{eq:temp_optimize} during training.  {See Table \ref{tab:hyperparams} in Appendix \ref{App:Hyperparameters} for more details of RL hyperparameters.}

\subsection{Reward design for optimal energy efficiency} \label{subsec:Reward}

We propose a hyperparameter-free reward function based on net power saving to discover energy-efficient flow control policies, calculated as the difference between the power saved from drag reduction $\Delta P_{D}$ and the power consumed from actuation $P_{act}$. Then, the power reward (``PowerR'') at the RL control frequency is
\begin{equation}
r_t= \underbrace{\Delta P_{D}}_{\textrm{power saved}}- \underbrace{P_{act}}_{\textrm{power spent}}.
\label{eq: PowerR}
\end{equation}
The power saved from drag reduction is given by 
\begin{equation}
\Delta P_{D} = P_{D0}-P_{Dt} = \left(\left\langle F_{D0}\right\rangle_T - \left\langle F_{Dt}\right\rangle_a\right) U_{\infty},
\label{eq: Drag Power Saving}
\end{equation}
where $P_{D0}$ is the time-averaged baseline drag power without control, and $\left\langle F_{D0}\right\rangle_T$ is the time-averaged baseline drag over a sufficiently long period. $P_{Dt}$ denotes the time-averaged drag power calculated from the time-averaged drag $\left\langle F_{Dt}\right\rangle_a$ during one RL step $t_a$. Specifically, $\langle ~ 
 \rangle_a$ quantities are calculated at each RL step using 125 DNS samples. 
The jet power consumption of actuation $P_{act}$ \citep{barros_bluff_2016} is defined as
\begin{equation}
P_{act}=\sum_{j=1}^2\left|\rho_{\infty} \langle U_{j} \rangle_a^{3} S_{j}\right| = \sum_{j=1}^2\left|\frac{ \left\langle a_{t}\right\rangle_a^{3}}{\rho_{\infty}^{2} S_{j}^{2}}\right|,
\label{eq: Actuation Power}
\end{equation}
where $\langle U_{j} \rangle_a$ is the average jet velocity, and $S_j$ denotes the area of one jet. 

The reward function given by  \req{eq: PowerR} quantifies the control efficiency of a controller directly. Thus, it guarantees the learning of a control strategy which simultaneously maximises the drag reduction and minimises the required control actuation. Additionally, this energy-based reward function avoids the effort of hyperparameter tuning.

All the cases in this work use the power-based reward function defined in \req{eq: PowerR} unless otherwise specified. For comparison, a reward function based on drag and lift coefficient (``ForceR'') is also implemented, as suggested by \citep{rabault_artificial_2019} with a pre-tuned hyperparameter $\epsilon=0.2$, as
\begin{equation}
r_t^a= C_{D0} - \left\langle C_{Dt}\right\rangle_a - \epsilon\left| \left\langle C_{Lt}\right\rangle_a\right|,
\label{eq: DragR}
\end{equation}
where $C_{D0}$ and $\left\langle C_{Dt}\right\rangle_a$ are calculated from a constant baseline drag and RL-step-averaged drag and lift. The RL-step-averaged lift $\left| \left\langle C_{Lt}\right\rangle_a\right|$ is used to penalise the amplitude of actuation on both sides of the body, avoiding excessive lift force (i.e. the lateral deflection of the wake reduces the drag but increases the side force), and indirectly penalising control actuation and the discovery of unrealistic control strategies. $\epsilon$ is a hyperparameter designed to balance the penalty on drag and lift force. 

The instantaneous versions of these two reward functions are also investigated for practical implementation purposes (both experimentally and numerically) because they can significantly reduce memory used during computation and also support a lower sampling rate. These instantaneous reward functions are computed only from observations at each RL step. In comparison, the reward functions above take into account the time history between two RL steps, while the instantaneous version of the power reward (``PowerInsR'') is defined as
\begin{equation}
r_{t,ins}= \Delta P_{D,ins}-P_{act,ins},
\label{eq: InsPowerR}
\end{equation}
where $\Delta P_{D,ins}$ is given by
\begin{equation}
\Delta P_{D,ins}= \left(\left\langle F_{D0}\right\rangle_T -  F_{Dt}\right) U_{\infty},
\label{eq: Ins Drag Power}
\end{equation}
and $P_{act,ins}$ is defined as 
\begin{equation}
P_{act,ins}=\sum_{j=1}^2\left|\rho_{\infty} \overline{U_{j}}^{3} S_{j}\right| = \sum_{j=1}^2\left|\frac{ a_{t}^{3}}{\rho_{\infty}^{2} S_{j}^{2}}\right|.
\label{eq: Ins Actuation Power}
\end{equation}
Notice that the definition of reward in \req{eq: InsPowerR} - \req{eq: Ins Actuation Power} is similar to \req{eq: PowerR} - \req{eq: Ins Actuation Power}, and the only difference is that the average operator $\langle ~ 
 \rangle_a$ is removed.
Similarly, the instantaneous version of the force-based reward function (``ForceInsR'') is defined as
\begin{equation}
r_{t,ins}^a= C_{D0} - C_{Dt} - \epsilon\left| C_{Lt}\right|.
\label{eq: InsDragR}
\end{equation}
In \S\ref{subsec:Rewards_Study}, we present results on the study of different reward functions and compare the RL performance.

\subsection{POMDP and dynamic  {feedback} controllers}\label{subsec:PM_Dynamic}

In practical applications, the Markov property \req{eq:markov_property} is often not valid due to noise, broken sensors, partial state information and delays. This means the observations available to the RL agent do not provide full or true state information, i.e. $o_t \neq s_t $, while in MDP $o_t = s_t $. Then, RL can be generalised as POMDP  defined as a tuple $(\mathcal{S}, \mathcal{A}, \mathcal{P}, \mathcal{R}, \mathcal{Y}, \mathcal{O})$, where $\mathcal{Y}$ is a finite set of observations $o_t$ and $\mathcal{O}$ is an observation function that relates observations to underlying states.

With only PM available in the flow environments (sensors on the  {downstream surface} of the body instead of in the wake), the spatial information is missing along the streamwise direction. Takens' embedding theorem \citep{takens_detecting_1981} states that the underlying dynamics of a high-dimensional dynamical system can be reconstructed from low-dimensional measurements with their time history. Therefore, past measurements can be incorporated into a sufficient statistic. Furthermore, convective delays may be introduced in the state observation since the sensors are not located in the wavemaker region of the flow. According to \cite{altman1992closed}, past actions are also required in the state of a delayed problem to reduce it into an undelayed problem. This is because a typical delayed-MDP (DMDP) implicitly subverts the Markov property, as the past measurements and actions only encapsulate partial information. 

Therefore, combining the ideas of augmenting past measurements and past actions, we form a sufficient statistic \citep{bertsekas_dynamic_2012} for reducing the POMDP problem to an MDP, defined as
\begin{equation}
I_k= [p_0,...,p_k, a_0,...,a_{k-1}],
\label{eq:Sufficient_statistic}
\end{equation}
which consists of the time history of pressure measurements $p_0,...,p_k$ and control actions $a_0,...,a_{k-1}$ at time steps $0,...,k$. This enlarged state at time $k$ contains all the information known to the controller at time $k$. 

However, the size of the sufficient statistic in \req{eq:Sufficient_statistic} grows over time, leading to a non-stationary closed-loop system and introducing a challenge in RL since  the number of inputs to the networks varies over time. This problem can be solved by reducing \req{eq:Sufficient_statistic} to a finite-history approximation \citep{white_iii_finite-memory_1994}. The controller using this finite-history approximation of the sufficient statistic is usually known as a ``finite-state'' controller, and the error of this approximation converges as the size of the finite history increases \citep{yu_near_2008}. The trade-off is that the dimension of the input increases based on the history length required.  
The nonlinear policy, which is parameterised by a neural network controller, has an algebraic description 
\begin{equation}
a_t \sim \pi_{\phi} \left(a_t \mid \underbrace{a_{t-1}, a_{t-2}, \ldots, a_{t-N_{fs}-1}}_{\text{past actions}}, p_t, \underbrace{p_{t-1}, p_{t-2}, \ldots, p_{t-N_{fs}}}_{\text{past measurements}} \right),
\label{eq:NARX}
\end{equation}
where $p_t$ represents pressure measurements at time step $t$, and $N_{fs}$ denotes the size of the finite history. The above expression \label{eq:NARX} is equivalent to a nonlinear autoregressive exogenous model (NARX). 

A ``frame stack'' technique is used to feed the ``finite history sufficient statistic'' to the RL agent as input to both the actor and critic neural networks. Frame stack constructs the observation $o_t$ from the latest actions and measurements at step $t$ as a ``frame'' $ o_t = (a_{t-1}, p_t)$, and piles up the finite history of $N_{fs}$ frames together into a stack. The number of stacked frames is equivalent to the size of the finite history $N_{fs}$. 

The neural network controller trained as a NARX model benefits from past information to approximate the next optimised control action since the policy has been parameterised as a nonlinear transfer function. Thus, a controller parameterised as a NARX model is denoted as a ``dynamic  {feedback}'' controller because the time history in the NARX model contains dynamic information of the system. Correspondingly, a controller fed with only the latest actions  {$a_{t-1}$} and  {current} measurements  {$p_t$} is denoted as a ``static  {feedback}'' controller  {because no past information from the system is fed into the controller}. 

\begin{figure}
  \centerline{\includegraphics[width=13cm]{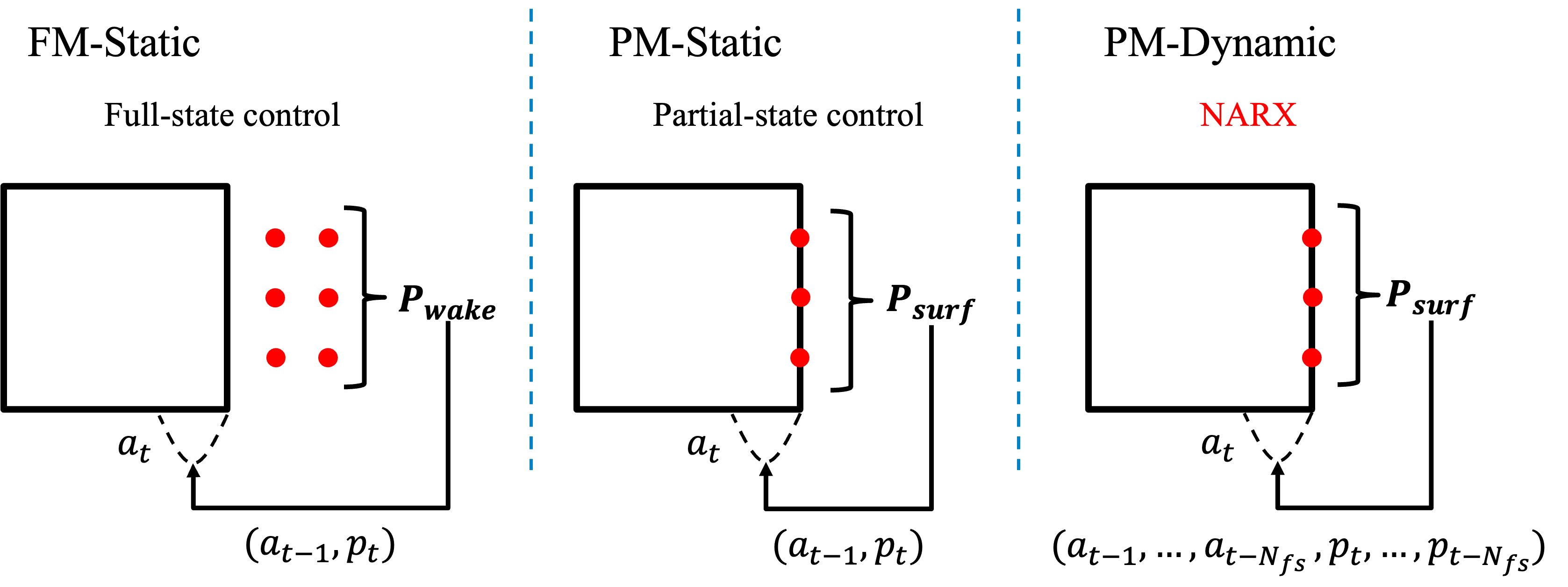}}
  \caption{Demonstration of a full-measurement (FM) environment with a static  {feedback} controller (``FM-Static''); a partial-measurement (PM) environment with a static  {feedback} controller (``PM-Static''); and a PM environment with a dynamic  {feedback} controller formulated as a NARX model (case ``PM-Dynamic''). The dashed curve represents the bottom blowing/suction jet, and the red dots demonstrate schematically the location of the sensors. }
\label{fig:Case_Demo}
\end{figure}

Figure \ref{fig:Case_Demo} demonstrates three cases with both FM and PM environments which will be investigated. In the FM environment, sensors are located in the wake as $\boldsymbol{P_{surf}}$ given by \req{eq:Probe_base}. In the PM environment, sensors are placed only on the back surface of the body as $\boldsymbol{P_{wake}}$  given by \req{eq:Probe_wake}. The static  {feedback} controller is employed in the FM environment, and both static and dynamic  {feedback} controllers are applied in the PM environment.
Results will be shown with $N_{fs} = 27$, and in \S\ref{subsec:Nfs}, a parametric study of the effect of the finite history length is presented.

%% file: Results.tex
\section{Results of RL active flow control}\label{sec:Results}

In this section, we discuss the converge of the RL algorithms for the three FM and PM cases (\S\ref{subsec:Convergence}) and evaluate their drag reduction performance (\S\ref{Result_drag_reduction}). A parametric analysis of the effect of NARX memory length is presented (\S\ref{subsec:Nfs}) and the isolated effect of including past actions as observations during the RL training and control (\S\ref{subsec:past_actions}). Studies of reward function (\S\ref{subsec:Rewards_Study}), sensor placement (\S\ref{subsec:Sensor_study}) and generalisability to Reynolds number changes (\S\ref{subsec:Res}) are presented, followed by a comparison of SAC and TQC algorithms (\S\ref{subsec:SACvsTQC}). 

\subsection{Convergence of learning}\label{subsec:Convergence}

We perform RL with the maximum entropy TQC algorithm to discover control policies for the three cases shown in figure \ref{fig:Case_Demo}, which maximise the net-power-saving reward function given by \req{eq: PowerR}. During the learning stage, each episode (1 DNS simulation) corresponds to $200$ non-dimensional time units.  To accelerate learning, $65$ environments run in parallel.

\begin{figure}
  \centerline{\includegraphics[width=12cm]{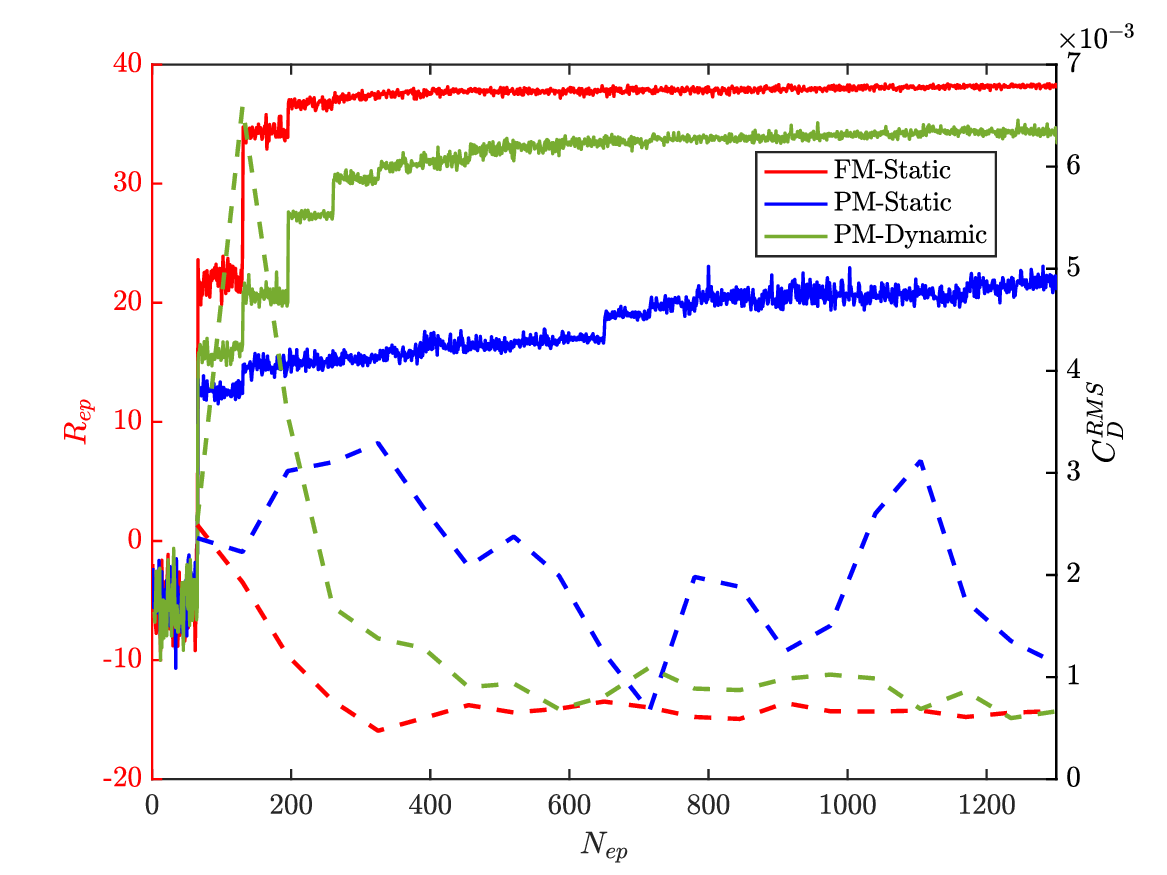}}
  \caption{Episode rewards (solid lines) and RMS of drag coefficient (dashed lines) against episode number during the maximum entropy reinforcement learning phase with TQC. }
\label{fig:Learning_Curve}
\end{figure}

Figure \ref{fig:Learning_Curve} shows the learning curves of the three cases.  Table \ref{tab:LearningConvergence} shows the number of episodes needed for convergence and relevant parameters for each case.
It can be observed from the curve of episode reward that the RL agent is updated after every 65 episodes, i.e. $1$ iteration, where the episode reward is defined as 
\begin{equation}
R_{ep} = \sum_{k=1}^{N_k} r_{k},
\label{eq:Epi_R}
\end{equation}
where $k$ denotes the $k^{th}$ RL step in one episode and $N_k$ is the total number of samples in one episode.
The root mean square (RMS) value of the drag coefficient, $C_D^{RMS}$, at the asymptotic regime of control, is also shown to demonstrate convergence, defined as 
$C_D^{RMS} = \sqrt { (\mathcal{D}(\langle C_D\rangle_{env}))^2 }$,
where the operator $\mathcal{D}$ detrends the signal with a $9^{th}$-order polynomial and removes the transient part, and $\langle ~ \rangle_{env}$ denotes the average value of parallel environments in a single iteration. 

\begin{table}
  \begin{center}
\def~{\hphantom{0}}
  \begin{tabular}{lcccccc}
    
      Environment  & Algorithm  &  $N_{c}$ & $R_{ep,c}$ & (Layers, Neurons) & $N_{fs}$ & Number of Inputs \\ 
       FM-Static   & TQC & $325$ & $37.72$ & (3,512) & $0$ & $64p_t+2a_{t-1}$\\
       PM-Static   & TQC & $1235$ & $21.87$ & (3,512) & $0$ & $64p_t+2a_{t-1}$\\
       PM-Dynamic  & TQC & $715$ & $34.35$ & (3,512) & $27$ & $N_{fs} (64p_t+2a_{t-1})$\\
  \end{tabular}
  \caption{Number of episodes $N_{c}$ required for RL convergence in different environments. The episode reward $R_{ep,c}$ at the convergence point, the configuration of NN and the dimension of inputs are presented for each case. $N_{fs}$ is the finite-horizon length of past actions-measurements.}
  \label{tab:LearningConvergence}
  \end{center}
\end{table}

In figure \ref{fig:Learning_Curve}, it can be noticed that in the FM environment, RL converges after approximately $325$ episodes ($5$ iterations) to a   {nearly} optimal policy using a static   {feedback} controller. As will be shown in \S\ref{Result_drag_reduction}, this policy is globally optimal since the vortex shedding is fully attenuated and the jets converge to zero mass flow actuation, thus recovering the unstable base flow and the minimum drag state.  However, with the same static   {feedback} controller in a PM environment (POMDP), the RL agent fails to discover the   {nearly} optimal solution, requiring around $1235$ episodes for convergence but only obtaining a relatively low episode reward.
Introducing a dynamic   {feedback} controller in the PM environment, the RL agent convergences to a near-optimal solution in 735 episodes. The dynamic   {feedback} controller trained by RL achieves a higher episode reward (34.35) than the static   {feedback} controller in the PM case (21.87), which is close to the FM case (37.72). The learning curves illustrate that using a finite horizon of past actions-measurements ($N_{fs} = 27$) to train a dynamic   {feedback} controller in the PM case improves learning in terms of speed of convergence and accumulated reward achieving nearly optimal performance with only wall pressure measurements.

\subsection{Drag reduction with dynamic RL controllers} \label{Result_drag_reduction}

\begin{figure}
  \centerline{\includegraphics[width=13cm]{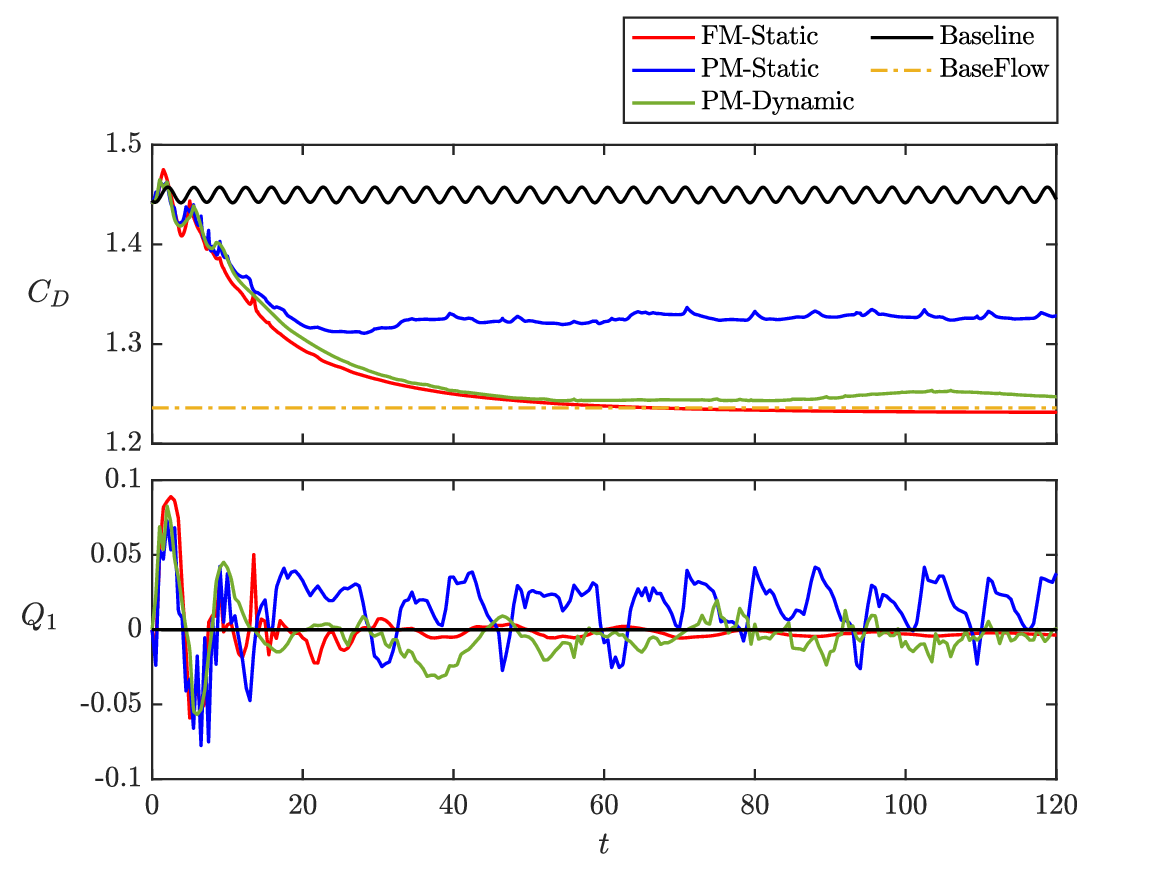}}
  \caption{Top figure: Drag coefficient $C_D$ without control (``Baseline'') and with active flow control by RL in both FM and PM cases. In PM cases, control results with a dynamic and static   {feedback} controller are presented. The dot-dashed line represents the base flow $C_{Db}$. Bottom figure: The mass flow rate $Q_1$ of one of the blowing and suction jets.}
\label{fig:TQC_FMPM}
\end{figure}

The trained controllers for the cases shown in figure \ref{fig:Case_Demo} are evaluated to obtain the results shown in figure \ref{fig:TQC_FMPM}.   {Evaluation tests are performed for 120 non-dimensional time units to show both transient and asymptotic dynamics of the closed-loop system.}
Control is applied at $t=0$ with the same initial condition for each case, i.e. steady vortex shedding with average drag coefficient $\langle C_{D0}\rangle \approx 1.45$ (baseline without control). Consistent with the learning curves, the difference in control performance in the three cases can be observed both from the drag coefficient $C_D$ and the actuation $Q_1$.
  {The drag reduction is quantified by a ratio $\eta$ using the asymptotic time-averaged drag coefficient with control $C_{Da} = \langle C_{D}\rangle_{t \in [80,120]}$, the drag coefficient $C_{Db}$ of the base flow (details presented in Appendix \ref{App:BaseFlow}), and the baseline time-averaged drag coefficient without control $\langle C_{D0}\rangle$, as
\begin{equation}
\eta = \frac{\langle C_{D0}\rangle - C_{Da}}{\langle C_{D0}\rangle - C_{Db}} \times 100\%.
\label{eq:drag_reduction}
\end{equation}}

\begin{itemize}

\item {\bf FM-Static:} With a static   {feedback} controller trained in a full-measurement environment, a drag reduction of $\eta = 101.96\%$ is obtained with respect to the base flow (steady unstable fixed point; maximum drag reduction). This indicates that an RL controller informed with full-state information can entirely stabilise the vortex shedding and cancel the unsteady part of the pressure drag.

\item {\bf PM-Static:} A static/memoryless controller in a partial-measurement environment leads to performance degradation and a drag reduction of   {$\eta = 56.00\%$} in the asymptotic control stage, i.e. after $t=80$, compared to the performance of ``FM-Static''. This performance loss can also be observed from the control actuation curve, as $Q_1$ oscillates with a relatively large fluctuation in ``PM-Static'' while it stays about zero in the ``FM-Static'' case. 
The discrepancy between FM and PM environments using a static   {feedback} controller reveals the challenge of designing a controller with a POMDP environment. The RL agent cannot fully identify the dominant dynamics with only partial measurements on the   {downstream} surface of the bluff body, resulting in sub-optimal control behaviour.

\item{\bf PM-Dynamic:} With a dynamic   {feedback} controller (NARX model presented in \S\ref{subsec:PM_Dynamic}) in a partial-measurement environment, the vortex shedding is stabilised and the dynamic   {feedback} controller achieves   {$\eta = 97.00\%$} of the maximum drag reduction after time $t=60$. Although there are minor fluctuations in the actuation $Q_1$, the energy spent in the synthetic jets is significantly lower compared to the ``PM-Static'' case. Thus, a dynamic   {feedback} controller in PM environments can achieve nearly optimal drag reduction, even if the RL agent only collects information from pressure sensors on the   {downstream} surface of the body. The improvement in control indicates that the POMDP due to the PM condition of the sensors can be reduced to an approximate MDP by training a dynamic   {feedback} controller with a finite horizon of past actions-measurements. Furthermore, high-frequency action oscillations, which can be amplified with static   {feedback} controllers, are attenuated in the case of dynamic   {feedback} control. These encouraging and unexpected results support the effectiveness and robustness of model-free RL control in practical flow control applications, in which sensors can only be placed on a solid surface/wall.

\end{itemize}

\begin{figure}
  \centerline{\includegraphics[width=12cm]{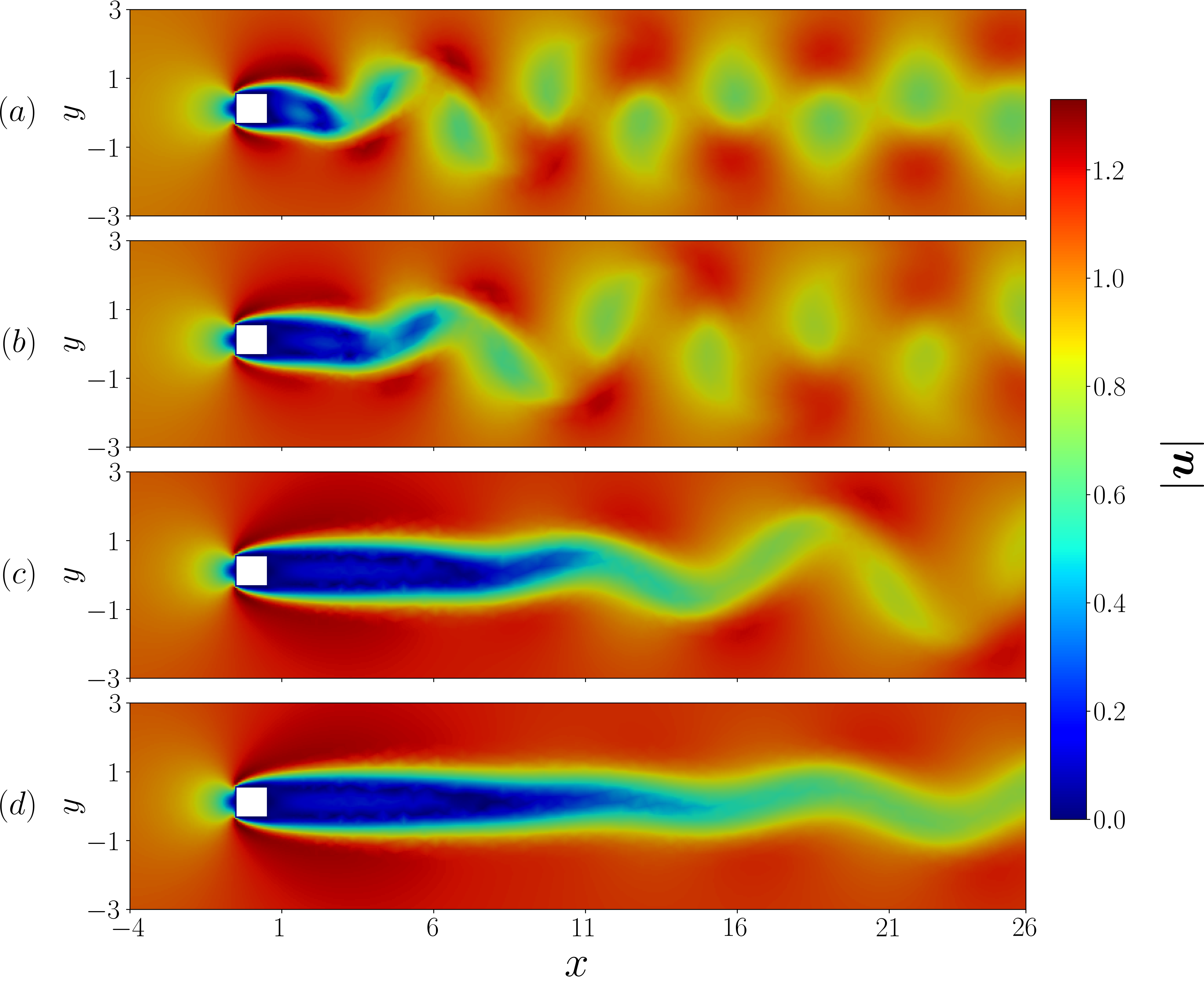}}
  \caption{Contours of velocity magnitude   {$|\boldsymbol{u}|$} in the asymptotic regime of control (at $t=100$).   {An area of $(-4,26)$ in the x direction and $(-3,3)$ in the y direction is presented for visualisation.} (a) ``Baseline'' (no control); (b) ``PM-Static''; (c) ``PM-Dynamic''; (d) ``FM-Static''.}
\label{fig:Contour}
\end{figure}

In figure \ref{fig:Contour}, snapshots of the velocity magnitude   {$|\boldsymbol{u}| = \sqrt{u^2+v^2}$} are presented for ``Baseline'' without control, ``PM-Static'', ``PM-Dynamic'' and ``FM-Static'' control cases. Snapshots are captured at $t=100$ in the asymptotic regime of control. A vortex-shedding structure of different strengths can be observed in the wake of all three controlled cases. In ``PM-Static'', the recirculation area is lengthened compared to the baseline flow, corresponding to base pressure recovery and pressure drag reduction. A longer recirculation area can be noticed in ``PM-Dynamic'' due to the enhanced attenuation of vortex shedding and pressure drag reduction. The dynamic   {feedback} controller in the PM case renders a $326.22\%$ increase of recirculation area with respect to the baseline flow, while only a $116.78\%$ increase is achieved by a static   {feedback} controller. The ``FM-Static'' case has the longest recirculation area, and the vortex shedding is almost fully stabilised, which is consistent with the drag reduction shown in figure \ref{fig:TQC_FMPM}.

\begin{figure}
\centering
\begin{subfigure}{0.46\textwidth}
    \includegraphics[width=\textwidth]{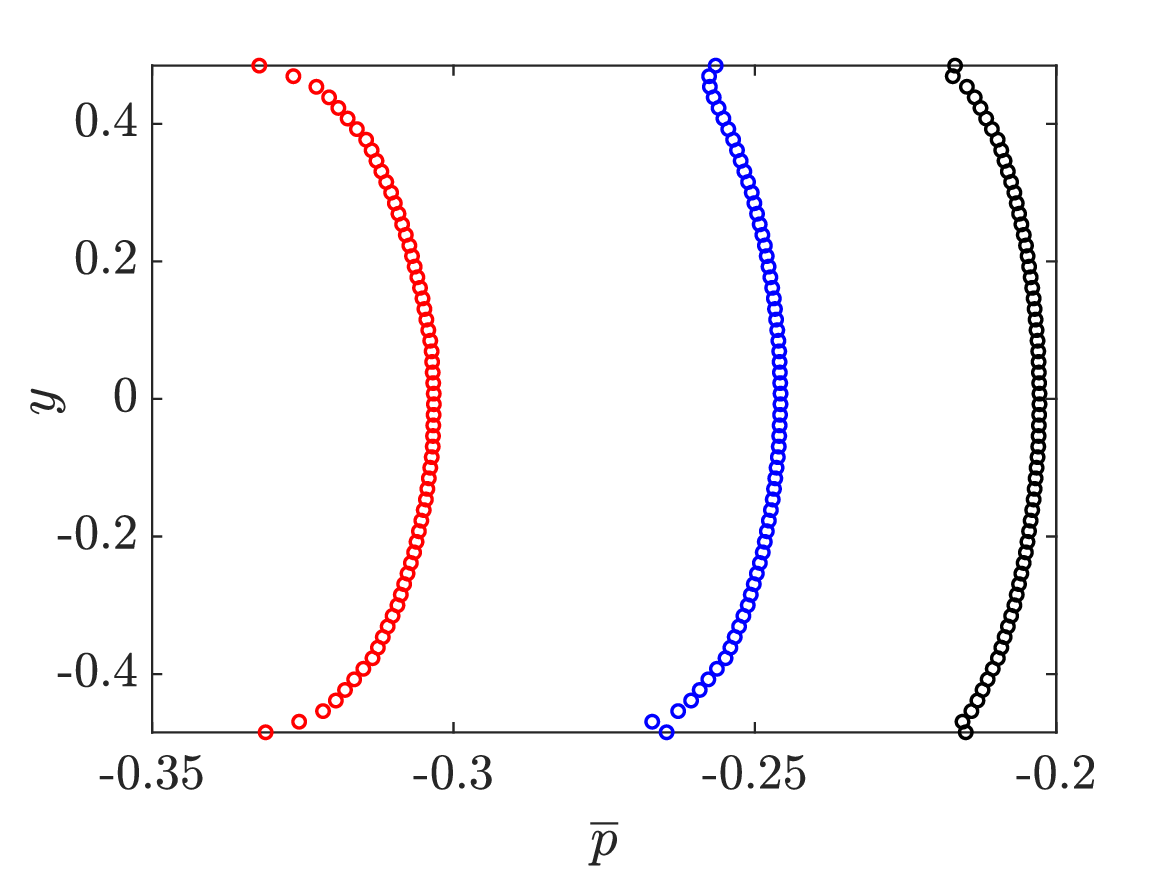}
    \caption{}
    \label{fig:first}
\end{subfigure}
\begin{subfigure}{0.46\textwidth}
    \includegraphics[width=\textwidth]{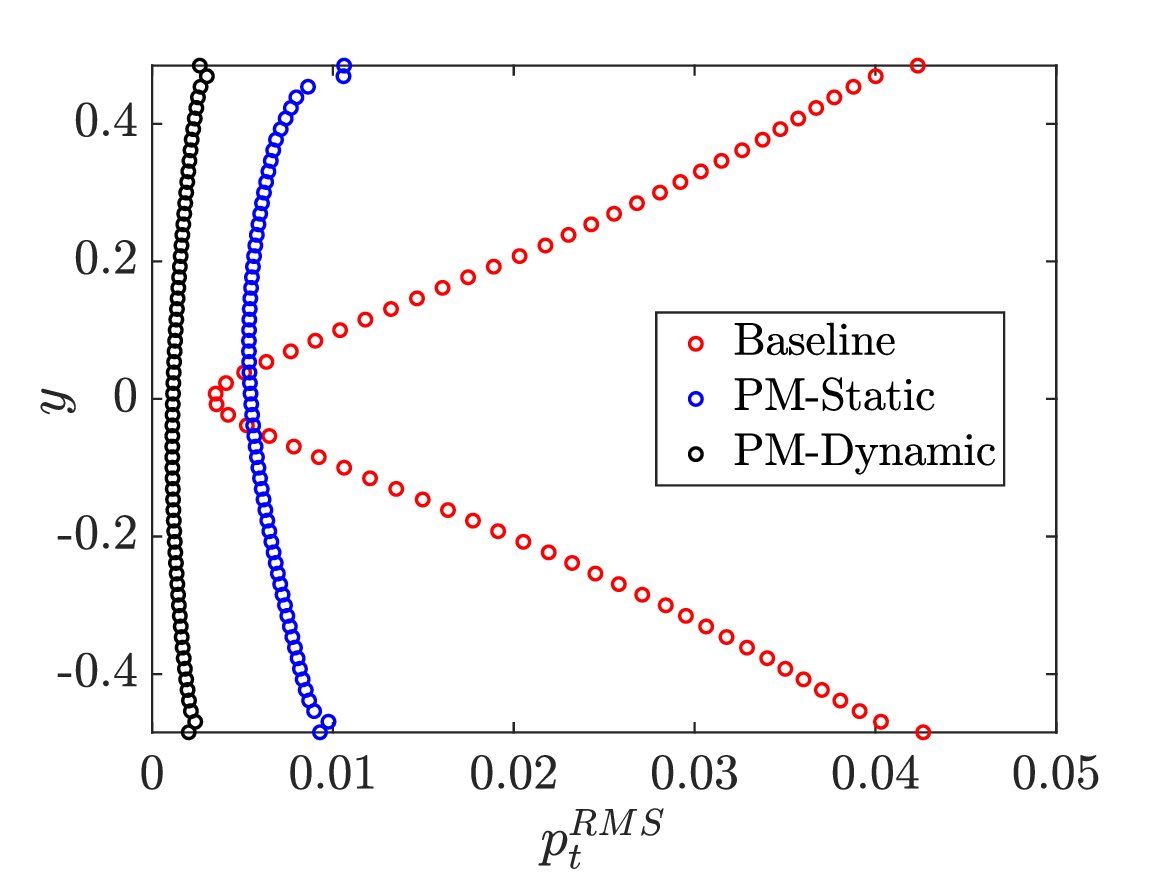}
    \caption{}
    \label{fig:second}
\end{subfigure}
        
\caption{Mean (a) and RMS (b) base pressure for controlled and uncontrolled cases from the $64$ wall sensors on the downstream surface of the bluff body base.}
\label{fig:Obs}
\end{figure}

Figure \ref{fig:Obs} presents first- and second-order base pressure statistics for the baseline case without control and PM cases with control. In figure \ref{fig:Obs}(a), the time-averaged value of base pressure, $\overline{p}$, demonstrates the base pressure recovery after control is applied. Due to flow separation and recirculation, the time-averaged base pressure is higher at the middle of the   {downstream surface}, which is retained with control. The base pressure increase is directly linked to pressure drag reduction, which quantifies the control performance of both static and dynamic   {feedback} controllers. Up to $49.56\%$ of pressure increase at the centre of the   {downstream surface}  is obtained in the ``PM-Dynamic'' case, while only $21.15\%$ can be achieved by a static   {feedback} controller. In figure \ref{fig:Obs}(b), the base pressure RMS is shown. For the baseline flow, strong vortex-induced fluctuations of the base pressure can be noticed around the top and bottom   {on the downstream surface} of the bluff body. In the ``PM-Static'' case, the RL controller   {partially suppresses} the vortex shedding, leading to a sub-optimal reduction of the pressure fluctuation. The sensors close to the top and bottom corners are also affected by the synthetic jets, which change the RMS trend for the two top and bottom measurements. In the ``PM-Dynamic'' case,  the pressure fluctuations are nearly zero for all the measurements on the   {downstream surface}, highlighting the success of vortex shedding suppression by a dynamic RL controller in a PM environment.

\begin{figure}
  \centerline{\includegraphics[width=13cm]{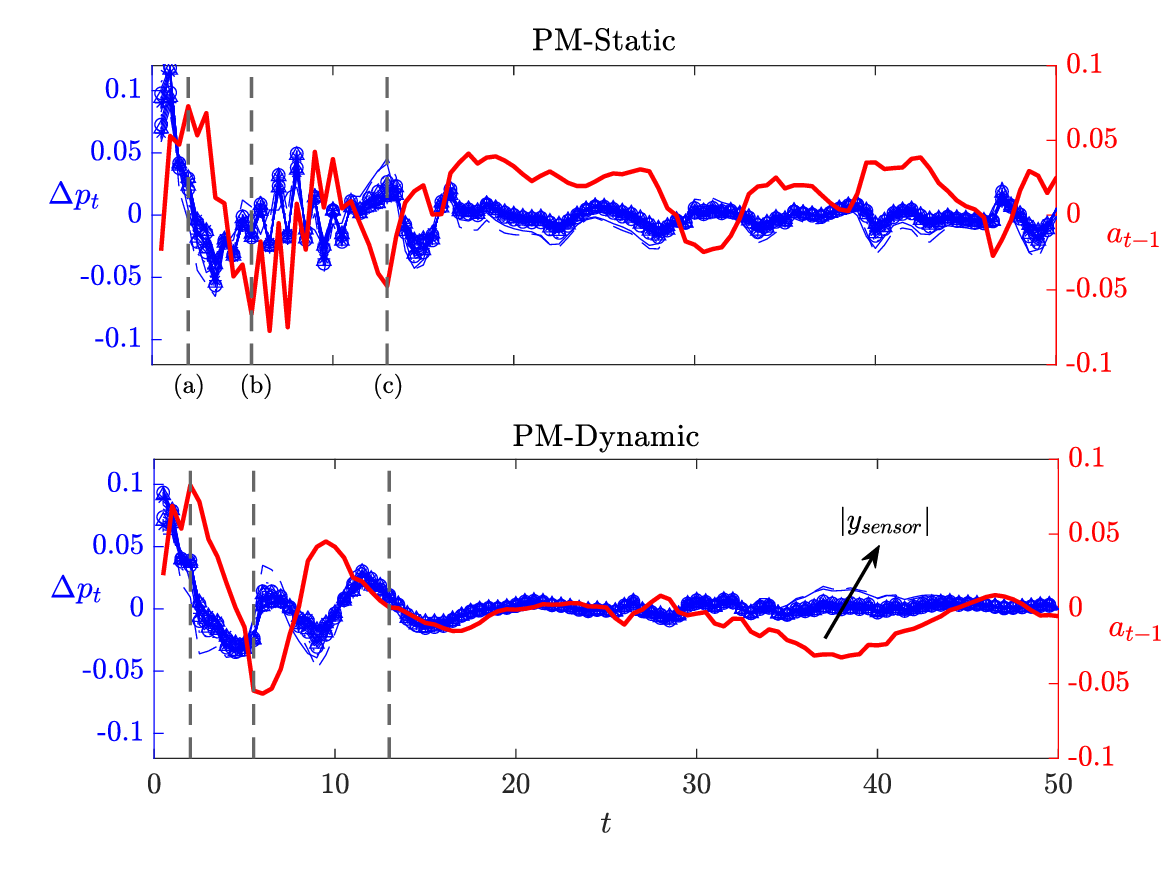}}
  \caption{  {Time series of pressure differences $\Delta p_t$ (blue) and action $a_{t-1}$ (red) for ``PM-Static'' (top) and ``PM-Dynamic'' (bottom) cases. Control is  applied at $t=0$. The arrows are pointing from low to high values of $|y_{sensor}|$ among $\Delta p_t$ curves. The vertical dashed lines mark the time instances of the vorticity snapshots in figure \ref{fig:ContourComparision}. }}
  
\label{fig:Action_analysis}
\end{figure}

The differences between static and dynamic controllers in PM environments are further elucidated in figure \ref{fig:Action_analysis} by examining  the time series of pressure differences $\Delta p_t$ from surface sensors (control input) and control actions $a_{t-1}$ (output). The pressure differences are calculated from sensor pairs at $y=\pm y_{sensor}$, where $y_{sensor}$ is defined in Eq. \req{eq:Probe_base}. For $N=64$, there are 32 time series of $\Delta p_t$ for each case. 
During  the initial stages of control ($t \in [0,11]$), the control actions are similar  for the two PM cases and they deviate for $t>11$, resulting in discernible control performance at the asymptotic regime. 
At the initial stages, the controllers operate in nearly anti-phase to $\Delta p_t$, in order to eliminate the antisymmetric pressure component due to vortex shedding. The inability of the static controller to have a frequency dependent amplitude (and phase), manifests as well through the amplification of high frequency noise. For $t>11$, the static feedback controller continues to operate in nearly anti-phase to the pressure difference, resulting in partial stabilisation of unsteadiness. However, the dynamic feedback controller adjusts its phase and amplitude significantly, which attenuates the antisymmetric fluctuation of base pressure and drives $\Delta p_t$ to near zero. 

\begin{figure}
    \centering
    \includegraphics[width=\textwidth]{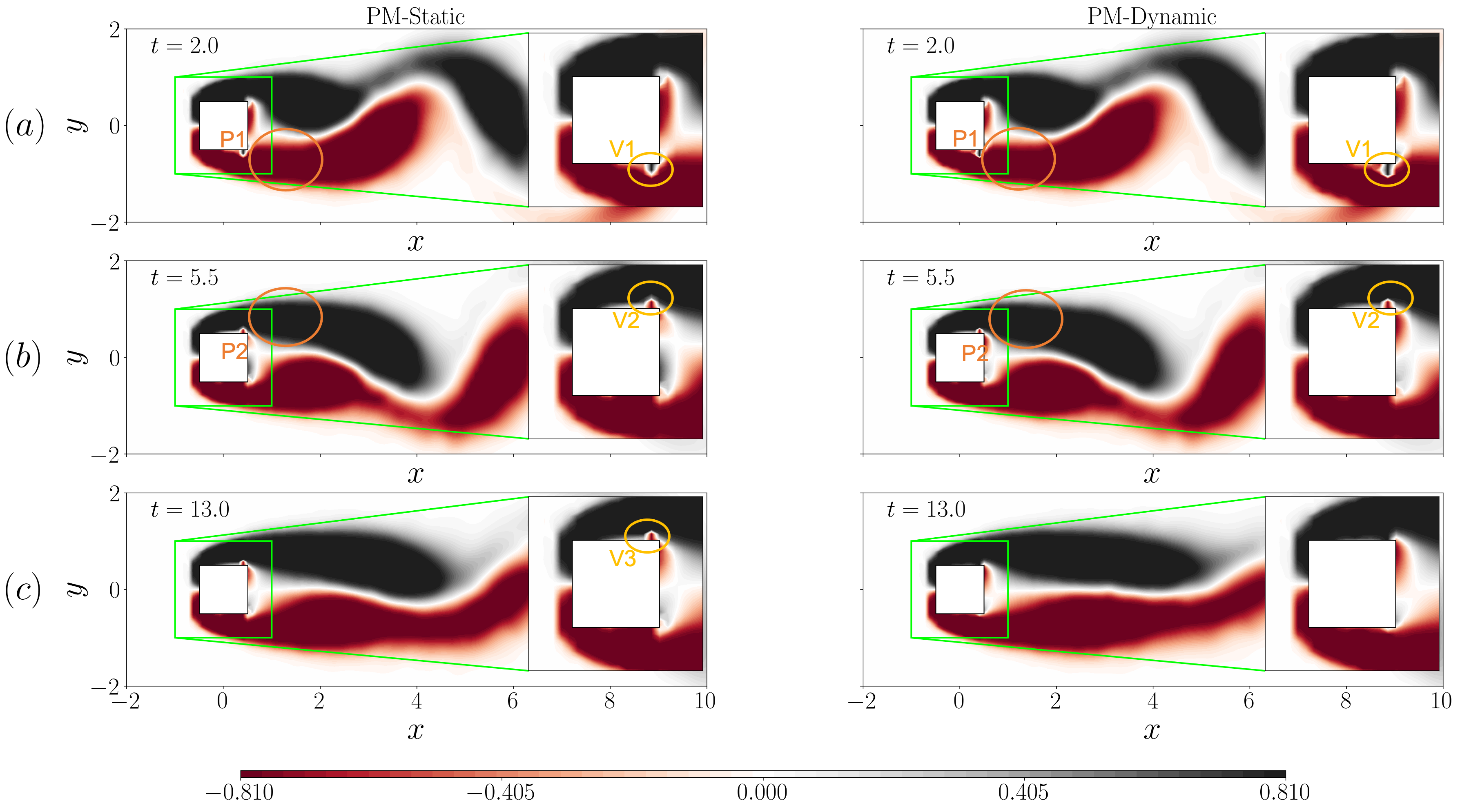}
    \caption{  {Vorticity snapshots at the transient phase of control. PM-Static (left); PM-Dynamic (right).}}
    \label{fig:ContourComparision}
\end{figure}

Figure \ref{fig:ContourComparision} shows instantaneous vorticity contours for PM-Dynamic and PM-Static cases, showing both the similarities and discrepancies between the two cases. At $t=2$, flow is expelled from the bottom jet for both cases, generating a clockwise vortex, termed V1. This V1 vortex, shown in black, works against the primary counter-clockwise vortex labelled as P1, depicted in red, emerging from the bottom surface. At $t=5.5$, a secondary vortex, V2, forms from the jets to oppose the primary vortex shedding from the top surface (labelled as P2). 
 At $t=13$, the suppression of the two primary vortices near the bluff body is evident in both cases, indicated by their less tilted shapes compared to the previous time instances. At $t=13$, the PM-Dynamic adjusted the phase of the control signal, which corresponds to a marginal action at this time instance at figure \ref{fig:Action_analysis}. Consequently, no additional counteracting vortex is formed in PM-Dynamic. However, in the PM-Static scenario, the jets generate a third vortex, labelled V3, which emerges from the top surface. This corresponds to a peak in the action of the PM-Static controller at this time. The inability of the PM-Static controller to adapt the amplitude/phase of the input/output behaviour results in suboptimal performance.

\subsection{Horizon of the finite-history sufficient statistic}\label{subsec:Nfs}

A parametric study on the horizon of the finite history in NARX (equation \req{eq:NARX}), i.e. the number of frames stacked $N_{fs}$, is presented in this section. Since the NARX model uses a finite horizon of past actions-measurements in  \req{eq:Sufficient_statistic}, the horizon of the finite history affects the convergence of the approximation \citep{yu_near_2008}. This approximation affects the optimisation during the learning of RL because it determines whether the RL agent can observe sufficient information to converge to an optimal policy. 

Since vortex shedding is the dominant instability to be controlled, the choice of $N_{fs}$ should intuitively link to the timescale of the vortex shedding period. The ``frames'' of observations are obtained every RL step ($0.5$ time units), while the vortex shedding period is $t_{vs}\approx6.85$ time units. Thus, $N_{fs}$ is rounded to integer values for different numbers of vortex shedding periods, as shown in table \ref{tab:Frame_Stack}.

\begin{figure}
  \centerline{\includegraphics[width=11cm]{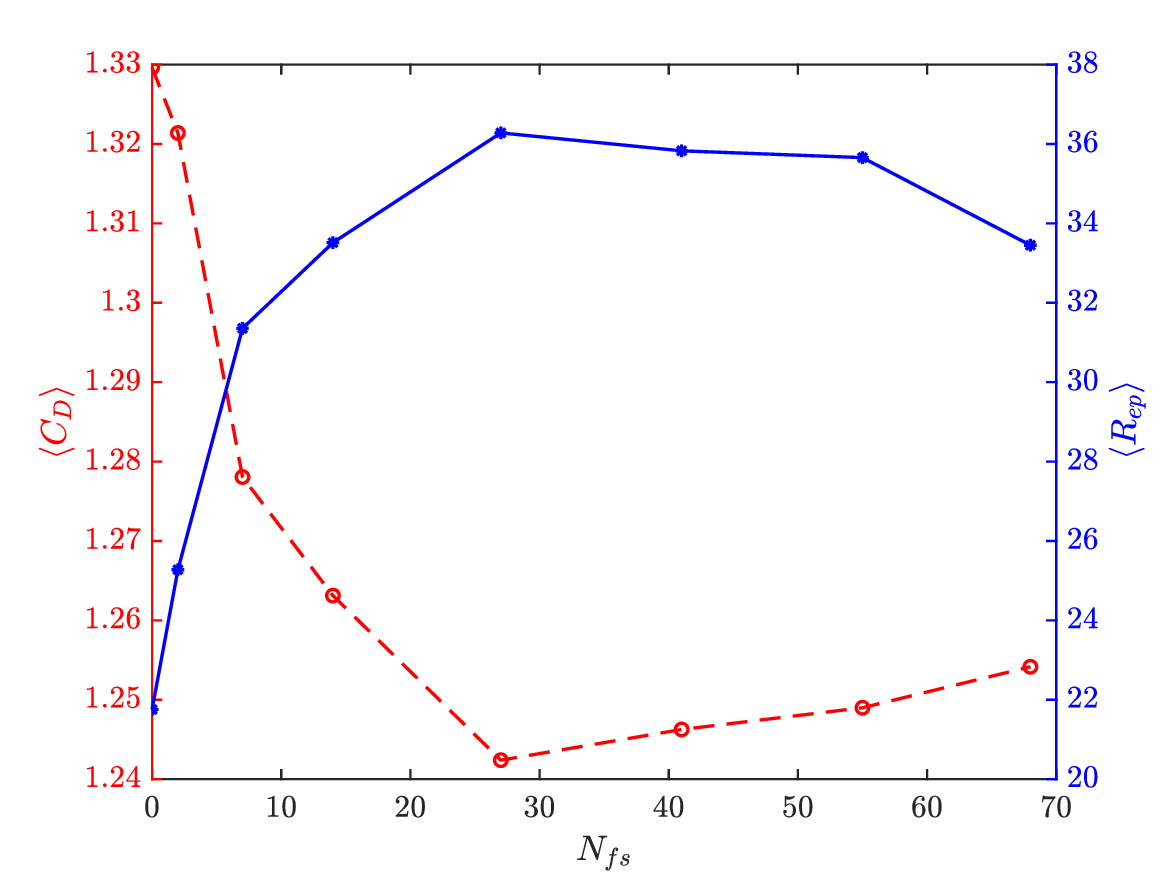}}
  \caption{Average drag coefficient $\langle C_{D}\rangle$ and average episode reward $\langle R_{ep}\rangle$ in PM cases against number history length (numbers of stacked frames) $N_{fs}$. $\langle C_{D}\rangle$ is obtained from the asymptotic regime of control. $\langle R_{ep}\rangle$ is calculated from 2 episodes after convergence of RL.}
  
\label{fig:Frame_Stack}
\end{figure}

\begin{table}
  \setlength{\tabcolsep}{12pt}
  \begin{center}
\def~{\hphantom{0}}
  \begin{tabular}{ccc}
      Number of  & Non-dimensional &  History length \\
      VS periods &    time units          &  ($N_{fs}$)         \\ [3pt]
      \hline
       0.5   & 3.43 & 7 \\
       1   & 6.85 & 14 \\
       2  & 13.70 & 27 \\
       3 & 20.55 & 41\\
       4 & 27.40 & 55\\
       5 & 34.25 & 68\\
  \end{tabular}
  \caption{Correspondence between the number of vortex shedding (VS) periods and frame stack (history) length in samples $N_{fs}$. The RL control step size is $t_a =0.5$, and $N_{fs}$ is rounded to an integer.}
  \label{tab:Frame_Stack}
  \end{center}
\end{table}

The results of time-averaged drag coefficients $\langle C_{D}\rangle$ after control and the average episode rewards $\langle R_{ep}\rangle$ in the final stage of training are presented in figure \ref{fig:Frame_Stack}. As $N_{fs}$ increases from 0 to 27, the performance of RL control improves, resulting in a lower $\langle C_{D}\rangle$ and a higher $\langle R_{ep}\rangle$. $N_{fs}=2$ is specially examined because the latent dimension of the vortex shedding limit cycle is 2. However, the control performance with $N_{fs}=2$ is marginally improved to the one with $N_{fs}=0$, i.e. a static   {feedback} controller. This result indicates that the horizon consistent with the vortex shedding dimension is not long enough for the finite horizon of past action measurements. The optimal history length to achieve stabilisation of the vortex shedding   {in PM environments} is 27 samples, which are equivalent to 13.5 convective time units or $\sim 2$ vortex shedding periods. 

With $N_{fs}=41$ and $N_{fs}=55$, the drag reduction and episode rewards drop slightly compared to $N_{fs}=27$. The decline in performance is non-negligible as $N_{fs}$ increases further to 68. This decline shows that excessive inputs to the neural networks (see table \ref{tab:LearningConvergence}), may impede training because more parameters need to be tuned or larger neural networks need to be trained. 

\subsection{Observation sequence with past actions}\label{subsec:past_actions}

Past actions (exogenous terms in NARX) facilitate reducing a POMDP to an MDP problem, as discussed in \S\ref{subsec:PM_Dynamic}. In the near-optimal control of a PM environment using a dynamic   {feedback} controller with inputs $\left( o_t, o_{t-1}, ..., o_{t-N_{fs}} \right)$, a sequence of observations $o_t = \left \{ p_t, a_{t-1}\right \}$ at step $t$ is constructed to include pressure measurements and actions. In the FM environment, due to the introduction of one-step delayed action due to the first-order-hold interpolation given by \req{eq:FOH_action}, the inclusion of the past action along with the current pressure measurement, meaning $o_t = \left \{ p_t, a_{t-1} \right \}$, is required even when the sensors are placed in the wake and cover the wavemaker region. 

\begin{figure}
  \centerline{\includegraphics[width=12cm]{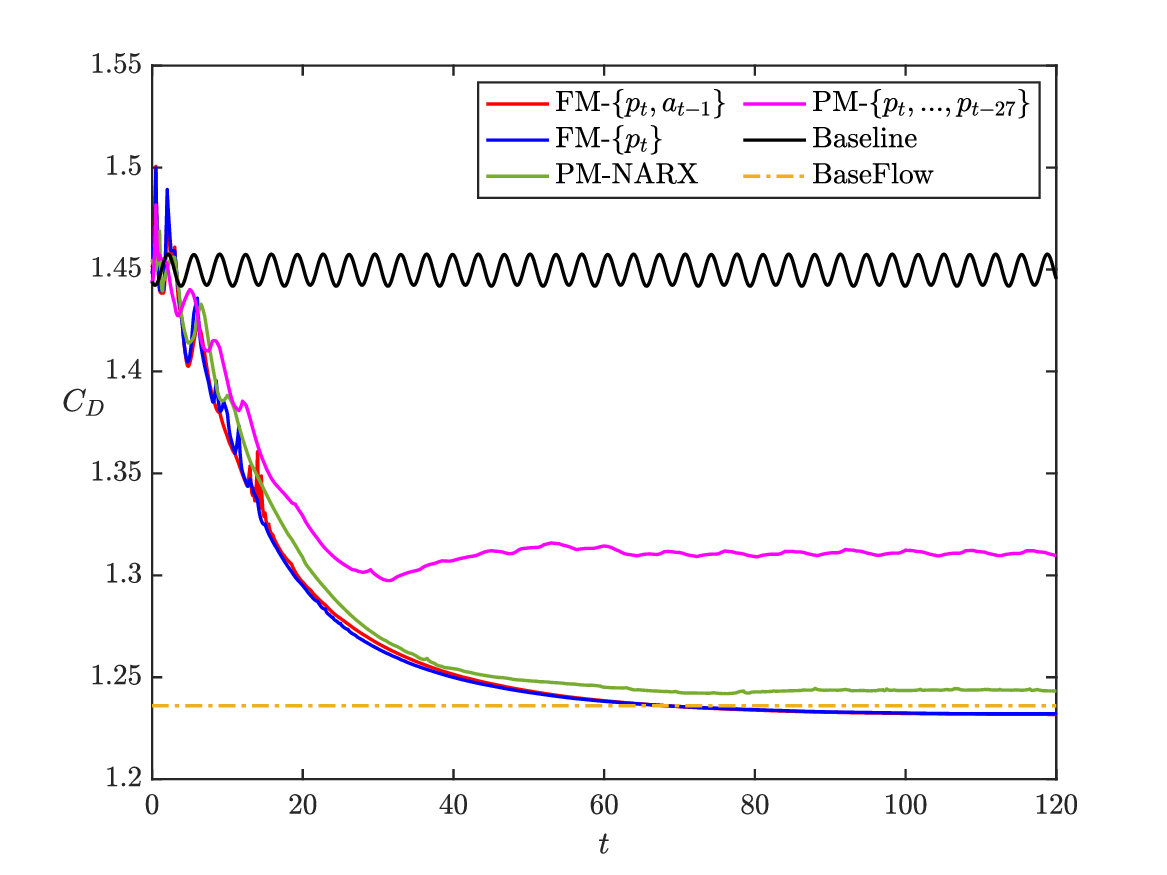}}
  \caption{Curves of drag coefficients after control being applied in both FM and PM environments. Results from FM cases are presented as references, while a performance difference can be observed in the PM cases with and without past actions included. }
\label{fig:ActionInObs}
\end{figure}

Figure \ref{fig:ActionInObs} presents the control performance for the same environment with and without past actions included.
In the FM case, there is no apparent difference between RL control with $o_t = \left \{ p_t, a_{t-1} \right \}$ or $o_t = \left \{ p_t \right \}$, which indicates that the inclusion of the past action is negligible to the performance. This is the case when the RL sampling frequency is sufficiently faster than the timescale of the vortex shedding dynamics. 
In PM cases, if exogenous action terms are not included in the observations but only the finite history of pressure measurements is used, the RL control fails to converge to a near-optimal policy, with only   {$\eta = 67.45\%$}  drag reduction. With past actions included, the drag reduction of the same environment increases up to   {$\eta = 97.00\%$}. 

The above results show that in PM environments, sufficient statistics cannot be constructed only from the finite history of measurements. Missing state information needs to be reconstructed by both state-related measurements and control actions. 

\subsection{Reward study}
\label{subsec:Rewards_Study}

In \S\ref{Result_drag_reduction}, a power-based reward function given by \req{eq: PowerR} has been implemented, and stabilising controllers can be learned by RL, as shown. In this section, RL control results with other forms of reward functions (introduced in \S\ref{subsec:Reward}) are provided and discussed.

\begin{figure}
  \centerline{\includegraphics[width=12cm]{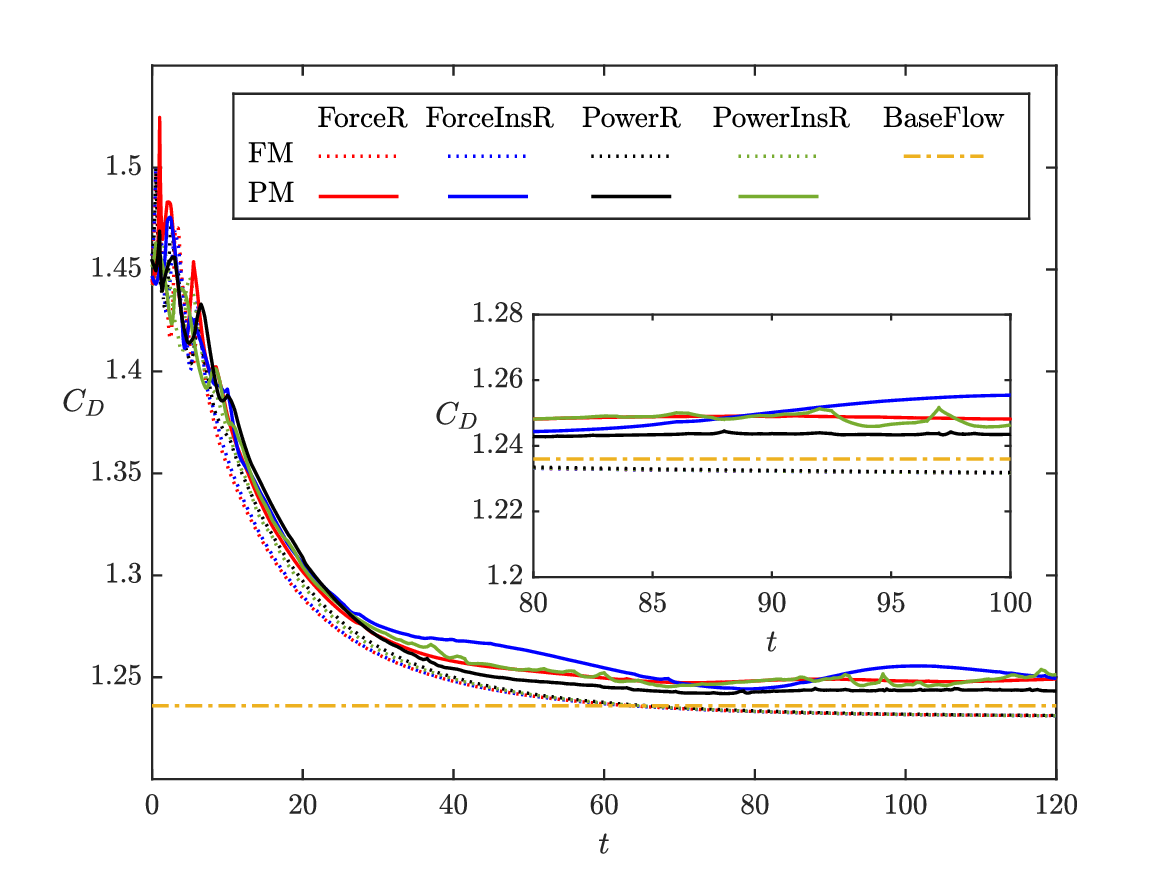}}
  \caption{Tests of RL-trained controllers with various reward functions. Drag coefficient $C_D$ curves are presented for each case. Dotted lines denote the cases with FM environments, while solid lines denote PM environments. The dash-dotted line represents $C_D$ in the base flow which has no vortex shedding. Control starts at $t=0$ with the same initial conditions for every case.}
\label{fig:Reward_Study}
\end{figure}

The control performance of RL control with the different reward functions is evaluated based on the drag coefficient $C_D$ shown in figure \ref{fig:Reward_Study}. Static   {feedback} controllers are trained in FM environments, and dynamic   {feedback} controllers are trained in PM environments. In FM cases, control performance is not sensitive to the choice of reward function (power or force-based).  
In PM cases, the discrepancies between RL-step time-averaged and instantaneous rewards can be observed in the asymptotic regime of control. The controllers with both rewards (power or force-based) achieve nearly optimal control performance, but there is some unsteadiness in the cases using instantaneous rewards due to slow statistical convergence of the rewards and limited correlation to the partial observations.

All four types of reward functions studied in this work achieve nearly optimal drag reduction around $100\%$. However, the energy-based reward (``PowerR'') offers an intuitive reward design, attributable to its physical properties and the dimensionally consistent addition of the constituent terms of the reward function. Further enhancing its practicality, since the power of the actuator can be directly measured, it avoids the necessity for hyperparameter tuning, as in the force-based reward. Additionally, the results show similar performance with both time-averaged between RL steps and instantaneous rewards, avoiding the necessity for faster sampling for the calculation of the rewards. This choice of reward function can be extended to various RL flow control problems and can be beneficial to experimental studies.

\subsection{Sensor configuration study with partial measurements}\label{subsec:Sensor_study}

\begin{figure}
  \centerline{\includegraphics[width=12cm]{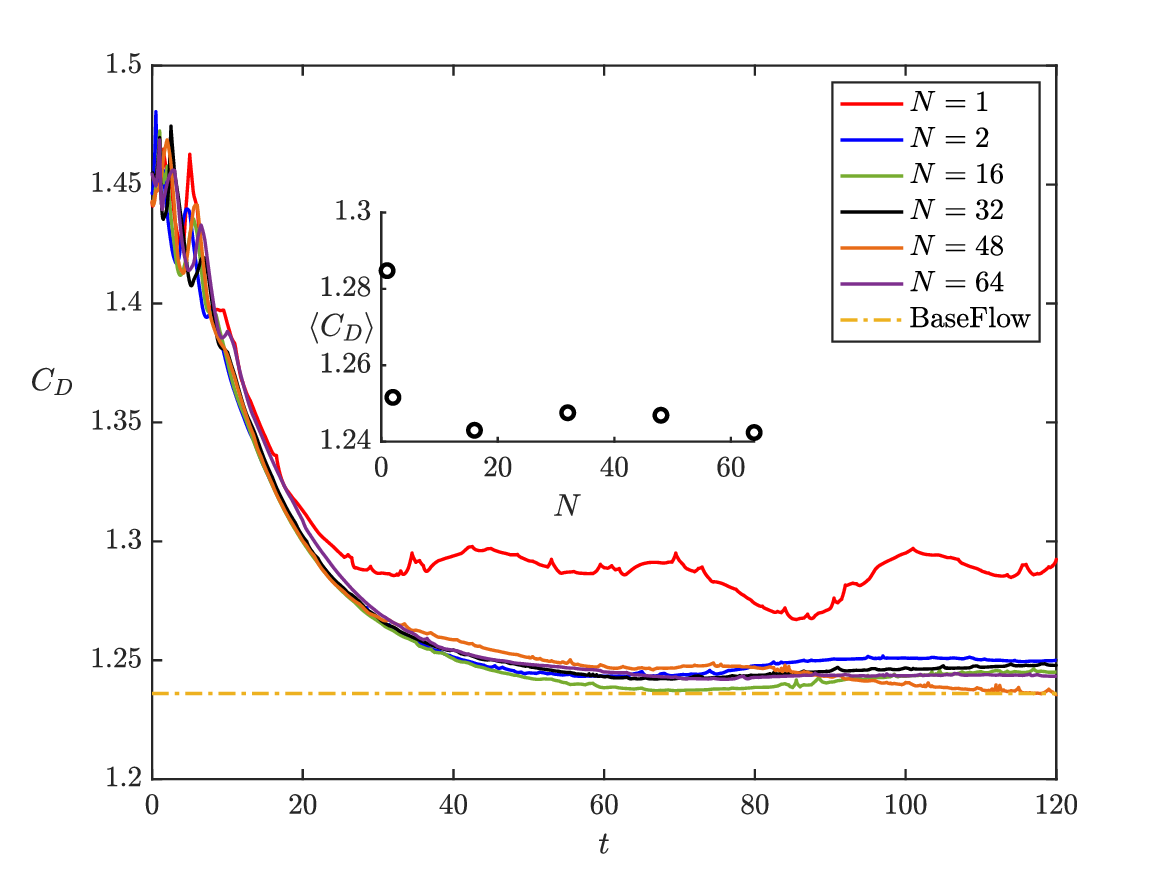}}
  \caption{  {Curves of drag coefficients after control applied at $t = 0$ in PM-Dynamic cases.
  Sensor configurations with different sensor numbers $N = 1, 2, 16, 32, 48, 64$ are tested. The dot-dashed line presents $C_D$ from the base flow. A sub-figure of asymptotic drag coefficient $\langle C_D \rangle$ (time-averaged value after $t = 80$) and probe number $N$ is presented.}}
\label{fig:Sensor_config}
\end{figure}

In the PM environment, the configuration of sensors (number and location on the downstream surface) may also affect the information contained in the observations and thus control performance. 
Control results of drag coefficient $C_D$ for different sensor configurations in PM-dynamic cases are presented in figure \ref{fig:Sensor_config}. In the configuration with $N = 2$, two sensors are placed at $y=\pm 0.25$, and for $N = 1$, only one sensor is placed at $y = 0.25$. Other configurations are consistent with equation \req{eq:Probe_base}. 

The $C_D$ curves in figure \ref{fig:Sensor_config} show that, as the number of sensors is reduced from 64 to 2, RL control achieves the same level of performance with minor discrepancies due to randomness in different learning cases. However, if RL control uses observations from only one sensor at $y = 0.25$, performance degradation can be observed in the asymptotic stage with 19.79\% on average less drag reduction. The sub-figure presents the relationship between the number of sensors and asymptotic drag coefficient $\langle C_D \rangle$. These results indicate a limit on sensor configuration for the use of the NARX-modeled controller to stabilise the vortex shedding. 

\begin{figure}
  \centerline{\includegraphics[width=12cm]{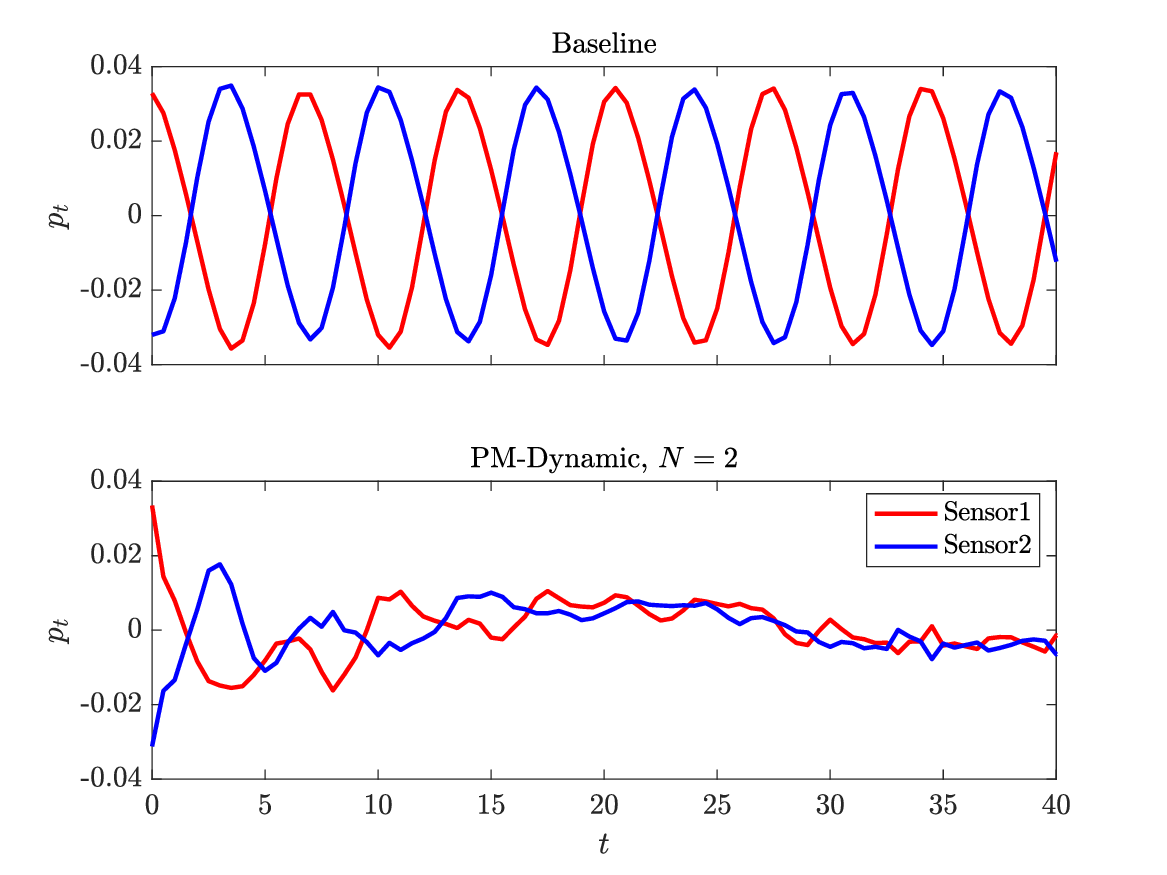}}
  \caption{  {Pressure measurements in $t\in[0,40]$ (early transient stage in the controlled case) from 2 surface sensors. ``Baseline'' without control (top); 
   ``PM-Dynamic'' with a NARX controller (bottom). All curves are detrended by a fifth-order polynomial to reveal the relationship between measurements from two sensors.}}
\label{fig:Pressure2Sensors}
\end{figure}

To understand the cause of performance degradation in the $N=1$ case, the pressure measurements from two sensors in both baseline and PM-Dynamic cases are presented in figure \ref{fig:Pressure2Sensors}. In the baseline case, two sensors are placed at the same location as the $N=2$ case ($y=\pm 0.25$) only for observations. It can be observed that the pressure measurements from two sensors are anti-symmetric since they are placed symmetrically on the downstream surface.
In the PM-Dynamic case, the NARX controller is used, and control is applied at $t=0$. In this closed-loop system, the anti-symmetric relationship between two sensors (from the symmetric position) is broken by the control actuation, and no correlation is evident. This can be seen during the transient dynamics, e.g. in $t \in [0,10]$. Therefore, when the number of sensors is reduced to $N=1$ by removing one sensor from the $N=2$ case, the dynamic feedback from the removed sensor cannot be fully reflected by the remaining sensor in the closed-loop system. This loss of information affects the fidelity of the control response to the dynamics of the sensor-removing side, causing suboptimal drag reduction in the $N=1$ scenario.

It should be noted that the configuration of 64 sensors is not necessary for control, as $N = 2$ or $N = 16$ also achieves nearly optimal performance. The number of sensors $N = 64$ in PM-Static environments is used for comparison with the FM-Static configuration (Eq. \ref{eq:Probe_wake}), which eliminates the effect from different input dimensions between two static cases. Also, 64 sensors sufficiently cover the downstream surface of the bluff body to avoid missing spatial information. 
The optimal configuration of sensors can be tuned with optimisation techniques such as \cite{paris_robust_2021}, but the results in figure \ref{fig:Sensor_config} indicate that RL adapts with nearly optimal performance to non-optimised sensor placement in the present environment.

\subsection{Performance of RL controllers to unseen $Re$} \label{subsec:Res}

\begin{figure}
  \centerline{\includegraphics[width=13cm]{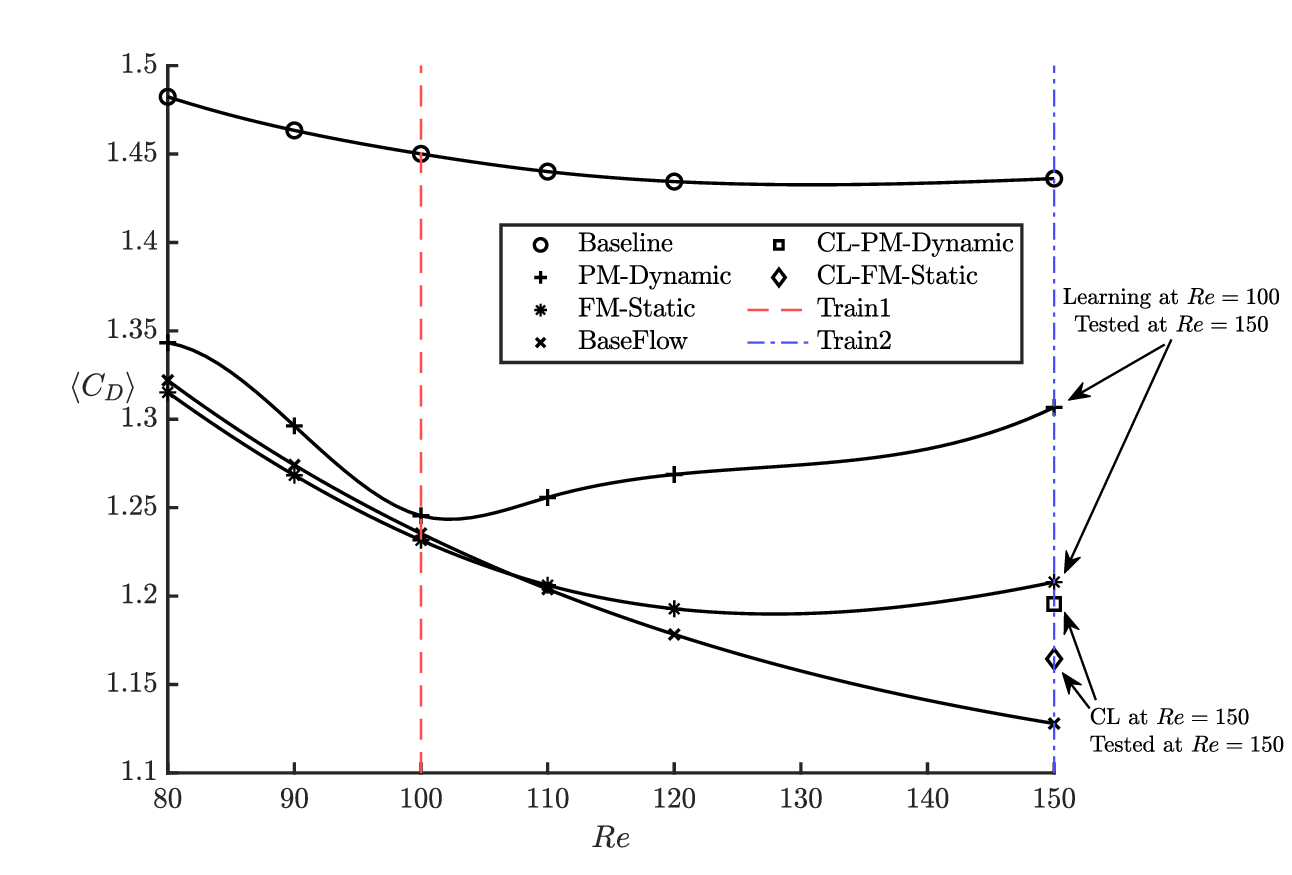}}
  \caption{Asymptotic drag coefficient $\langle C_D \rangle$ for ``Baseline'', ``Baseflow'', and tests of RL-trained controllers in both FM and PM environments with different $Re$. The controllers were trained at $Re=100$ (dashed line), and tested at $Re= 80, 90, 100, 110, 120, 150$. The controllers were trained again at $Re=150$ (dash-dotted line) and tested at $Re=150$ (square and diamond markers). All curves are fitted using a third-order spline.}
\label{fig:Res}
\end{figure}

The RL controller is tested at different Reynolds numbers, in order to examine its generalisability to environment changes. The controllers have been trained at $Re=100$ with both FM and PM conditions, and tested at $Re= 80, 90, 100, 110, 120, 150$. The controllers were further trained at $Re=150$, denoted as continual learning (CL), and tested again at $Re=150$. 

As shown in figure \ref{fig:Res}, in both ``PM-Dynamic'' and ``FM-Static'' cases, the RL controllers are able to reduce drag by $\eta=64.68\%$ in the worst case, when $Re$ is close to the training point at $Re=100$, i.e. the test cases with $Re= 80, 90, 100, 110, 120$. 
However, when applying the controllers trained at $Re=100$ to an environment at $Re=150$, the drag reduction drops to $\eta=41.98\%$ and $\eta = 74.04\%$ in PM-Dynamic and FM-Static cases, respectively.

Performing CL at $Re=150$, the drag reduction is improved to $\eta = 78.07\%$ in PM-Dynamic after 1105 training episodes while $\eta = 88.13\%$ in FM-Static after 390 episodes, with the same RL parameters as the training at $Re=100$.
Overall, the results of these tests indicate that the RL-trained controllers can achieve significant drag reduction in the vicinity of the training point (i.e. $\pm\%20$ $Re$ change). If the test point is far from the training point, a CL procedure can be implemented to achieve nearly optimal control.

\subsection{TQC vs SAC}\label{subsec:SACvsTQC}

\begin{figure}
  \centerline{\includegraphics[width=12cm]{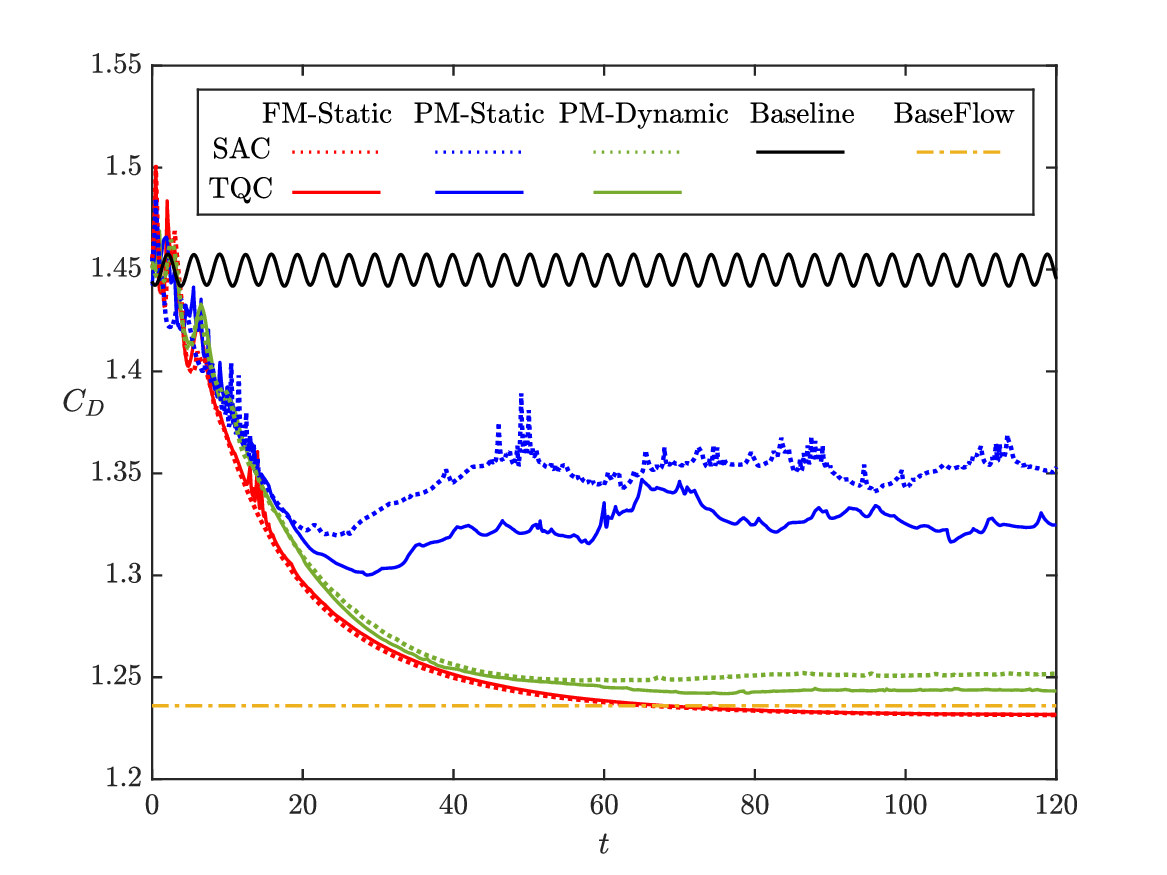}}
  \caption{Comparison of control performance in terms of $C_D$ between SAC and TQC. Control starts at $t=0$. Solid curves show the cases using TQC and ``Baseline'' while dotted curves show SAC. The dash-dotted curve corresponds to the baseflow  $C_D$.}
\label{fig:TQCvsSAC}
\end{figure}

Control results with TQC and SAC are presented in figure \ref{fig:TQCvsSAC} in terms of $C_D$. TQC shows a more robust control performance. In the case of FM, SAC might demonstrate a slightly more stable transient behaviour attributed to the fact that the quantile regression process in TQC introduced complexity to the optimisation process. Both controllers achieved an identical level of drag reduction in the FM case. 

However, in the context of the PM cases, it is observed that TQC outperforms SAC in drag reduction with both static and dynamic   {feedback} controllers. For static   {feedback} control, TQC achieved an average drag reduction of   {$\eta = 56.00\%$}, compared to the   {$\eta = 46.31\%$}  reduction achieved by SAC. The performance under dynamic   {feedback} control conditions is more compelling, where TQC fully reduced the drag, achieving   {$\eta = 97.00\%$}  of drag reduction, reverting it to a near-base-flow scenario. In contrast, SAC managed to achieve an average drag reduction of   {$\eta = 96.52\%$}.

The fundamental mechanism for updating Q-functions in RL involves selecting the maximum expected Q-functions among possible future actions. This process, however, can potentially lead to overestimation of certain Q-functions \citep{hasselt_double_2010}. In POMDP, this overestimation bias might be exacerbated due to the inherent uncertainty arising from the partial-state information. Therefore, the Q-learning-based algorithm, when applied to POMDPs, might be more prone to choosing these overestimated values, thereby affecting the overall learning and decision-making process.

As mentioned in \S\ref{subsec:SACTQC}, the core benefit of TQC under these conditions can be attributed to its advanced handling of the overestimation bias of rewards. By constructing a more accurate representation of possible returns, TQC provides a more accurate Q-function approximation than SAC. This process of modulating the probability distribution of the Q-function assists TQC in managing the uncertainties inherent in environments with only partial-state information. In this case, TQC can adapt more robustly to changes and uncertainties, leading to better performance in both static and dynamic feedback control tasks.

%% file: Conclusion.tex
\section{Conclusions}\label{sec:Conclusions}

In this study, maximum entropy RL with TQC has been performed in an active flow control application with partial measurements to learn a feedback controller for bluff body drag reduction. Neural network controllers have been trained by the RL algorithm to discover a drag reduction control strategy behind a 2D square bluff body at $Re=100$. By comparing the control performances in FM environments to PM environments, we showed a non-negligible degradation of RL control performance if the controller is not trained with full-state information. To solve this issue, we proposed a method to train a dynamic neural network controller with an approximation of a finite-history sufficient statistic and formulate the dynamic controller as a NARX model. The dynamic controller was able to improve the drag reduction performance in PM environments and achieve near-optimal performance ({drag reduction ratio $\eta = 97\% $} with respect to the baseflow drag) compared to a static controller ({$\eta = 56\% $}). We found that the optimal horizon of the finite history in NARX is approximately two vortex shedding periods when the sensors are located only on the base of the body. The importance of including exogenous action terms in the observations of RL is discussed, by pointing out the degradation of    {$\eta = 29.55\% $} on drag reduction if only past measurements are used in the PM environment. Also, we proposed a net power consumption design for the reward function based on the drag power savings and the power of the actuator. This power-based reward function offers an intuitive understanding of the closed-loop performance, whereas electromechanical losses can also be added directly, once a specific actuator is chosen. Moreover, its inherent feature of being hyperparameter-free contributes to a straightforward reward function design process in the context of flow control problems.    {Results from SAC are compared with TQC, and we showed the improvement by TQC, which attenuates overestimation in neural networks.}

It was shown that model-free RL was able to discover a nearly optimal control strategy without any prior knowledge of the system dynamics using partial realistic measurements, exploiting only input-output data from the simulation environment. Therefore, this particular study on RL-based active flow control in 2D laminar flow simulations can be seen as a    {step towards} controlling the complex dynamics of 3D turbulent flows    {in practical applications} by replacing the simulation environment with the experimental setup.    {Also, the frame stack method employed here to convert the POMDP to an MDP can be replaced by recurrent neural networks and attention-based architectures, which may further improve control performance in a scenario with complex dynamics.}

\

{\small
\noindent {\bf Funding.} We acknowledge support from the UKRI AI for Net Zero grant EP/Y005619/1.

\

\noindent  {\bf Data availability statement.} The open-source code of this project is available on the GitHub repository: https://github.com/RigasLab/Square2DFlowControlDRL-PM-NARX-SB3.
The code is developed from the work by \citet{rabault_artificial_2019} and \citet{rabault_accelerating_2019}, using a simulation environment by FEniCS v2017.2.0 \citep{logg_dolfin_2012}.
The RL algorithm is adapted to a version with SAC/TQC, implemented by Stable-Baselines3 and Stable-Baselines3-contrib \citep{stable-baselines3} in a PyTorch \citep{paszke2019pytorch} environment. See Appendix \ref{App:Sim_details} and the GitHub repository for more details of the simulation.

\

\noindent  {\bf Declaration of interests.} The authors report no conflict of interest.
}

%% file: Appendix.tex
\appendix

\section{Details of simulation environment} \label{App:Sim_details}

The simulation environment for solving the governing Navier-Stokes equations is adapted from \citet{rabault_artificial_2019} to the flow past a square bluff body. The boundary condition at the inflow boundary $\Gamma_I$ is set as a uniform velocity profile, and a zero-pressure condition is used at the outflow boundary $\Gamma_O$. Free-stream condition is used at the top and bottom boundary $\Gamma_D$ of the domain. The boundary on the bluff body is separated into body surface $\Gamma_W$ and jet area $\Gamma_j$, with a non-slip boundary condition and jet velocity profile, respectively. All the boundary conditions are formulated as
\begin{equation}
\begin{aligned}
u & = && U_{\infty} & \text { on } &\Gamma_I, \\
p & = && 0 & \text { on } &\Gamma_O, \\
u & = && U_{\infty} & \text { on } &\Gamma_D, \\
u & = && 0 & \text { on } &\Gamma_W, \\
\boldsymbol{u} & = && \boldsymbol{U_{j}} & \text { on } &\Gamma_j, j=1,2.
\end{aligned}
\label{eq:BCs}
\end{equation}

\begin{figure}
 \centerline{\includegraphics[width=13cm]{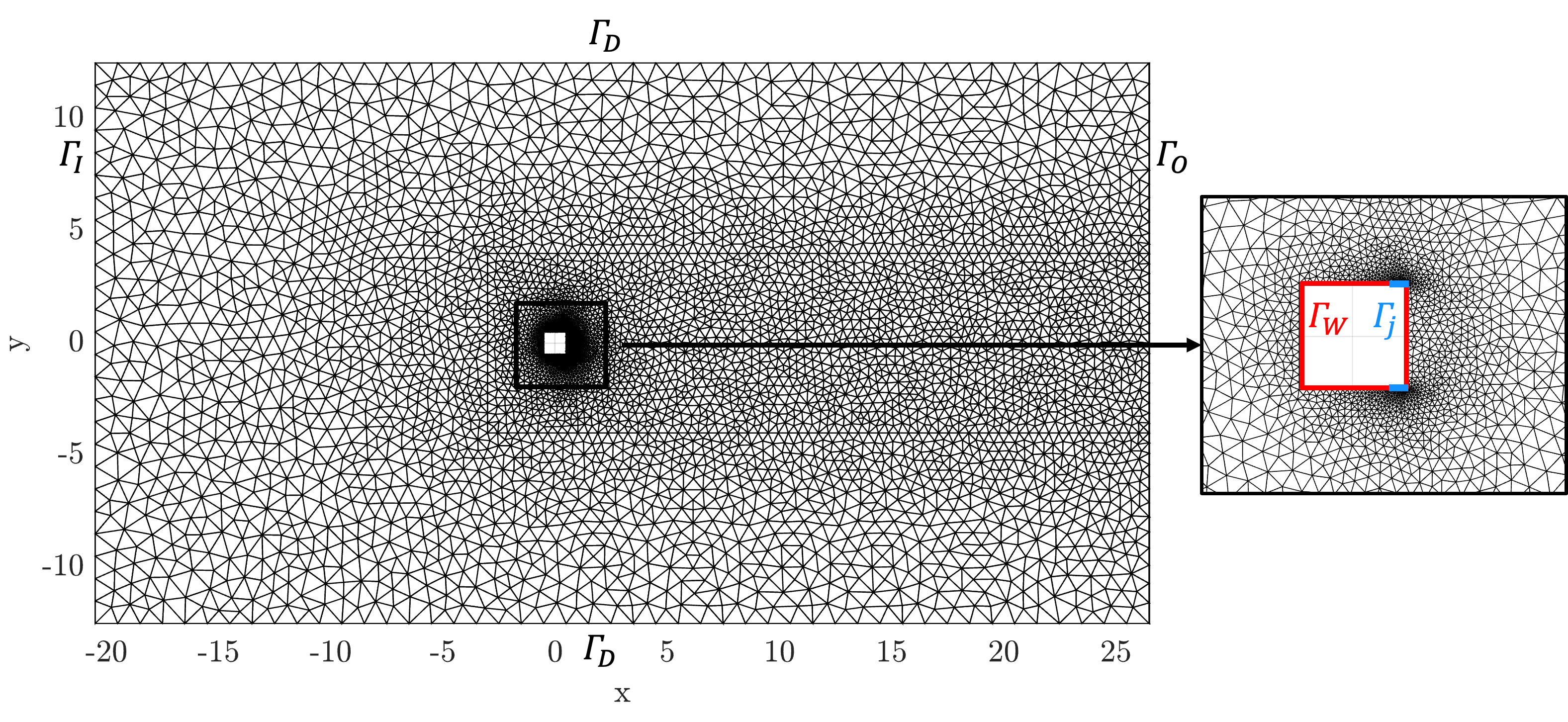}}
 \caption{{Computational mesh of the simulation domain; $x \in (-20.5,26.5)$ and $y \in (-12.5, 12.5)$. A zoom-in view around the bluff body is presented in the black rectangle at the right. Boundaries of the simulation domain, bluff body surface and jet area are denoted.}}
\label{fig:Mesh}
\end{figure}

The mesh of the simulation domain and a zoom-in view of the mesh around the square bluff body are presented in Fig. \ref{fig:Mesh}. The mesh is refined in the wake region with a ratio of 0.45 and near the body wall with a ratio of 0.075, with respect to the mesh size of the far field. Near the jet area, the mesh is further refined with a ratio of 0.015. More details can be found in the source code (see GitHub repository).

\section{Hyperparameters of RL} \label{App:Hyperparameters}

The RL hyperparameters to reproduce the result section (\S\ref{sec:Results}) are listed in table \ref{tab:hyperparams}.
\begin{table}
  \begin{center}
\def~{\hphantom{0}}
\begin{tabular}{l|l}
Hyperparameter & Value \\
\hline optimiser & Adam \\
learning rate & $10^{-4}$ \\
discount $(\gamma)$ & 0.99 \\
replay buffer size & $10^5$ \\
number of hidden layers (both actor and critic) & 3 \\
number of hidden units per layer & 512 \\
number of samples per minibatch & 128 \\
entropy target & $-\operatorname{dim}(\mathcal{A})$ \\
activation & $\operatorname{ReLU}$ \\
target smoothing coefficient $(\tau)$ & $5 \cdot 10^{-3}$ \\
target update interval & 1 \\
gradient steps & 48 \\
top quantiles to drop per net & 2 \\
number of quantiles per net & 25 \\

\end{tabular}
  \caption{   {Hyperparameters used by default in TQC. For SAC, ``top quantiles to drop per net'' is not used, and other parameters remain the same. For the entropy target, $-\operatorname{dim}(\mathcal{A})$ denotes the dimension of action space $\mathcal{A}$.}}
  \label{tab:hyperparams}
  \end{center}
\end{table}

\section{A long-run test of RL-trained controller} \label{App:LongEva}

\begin{figure}
  \centerline{\includegraphics[width=11cm]{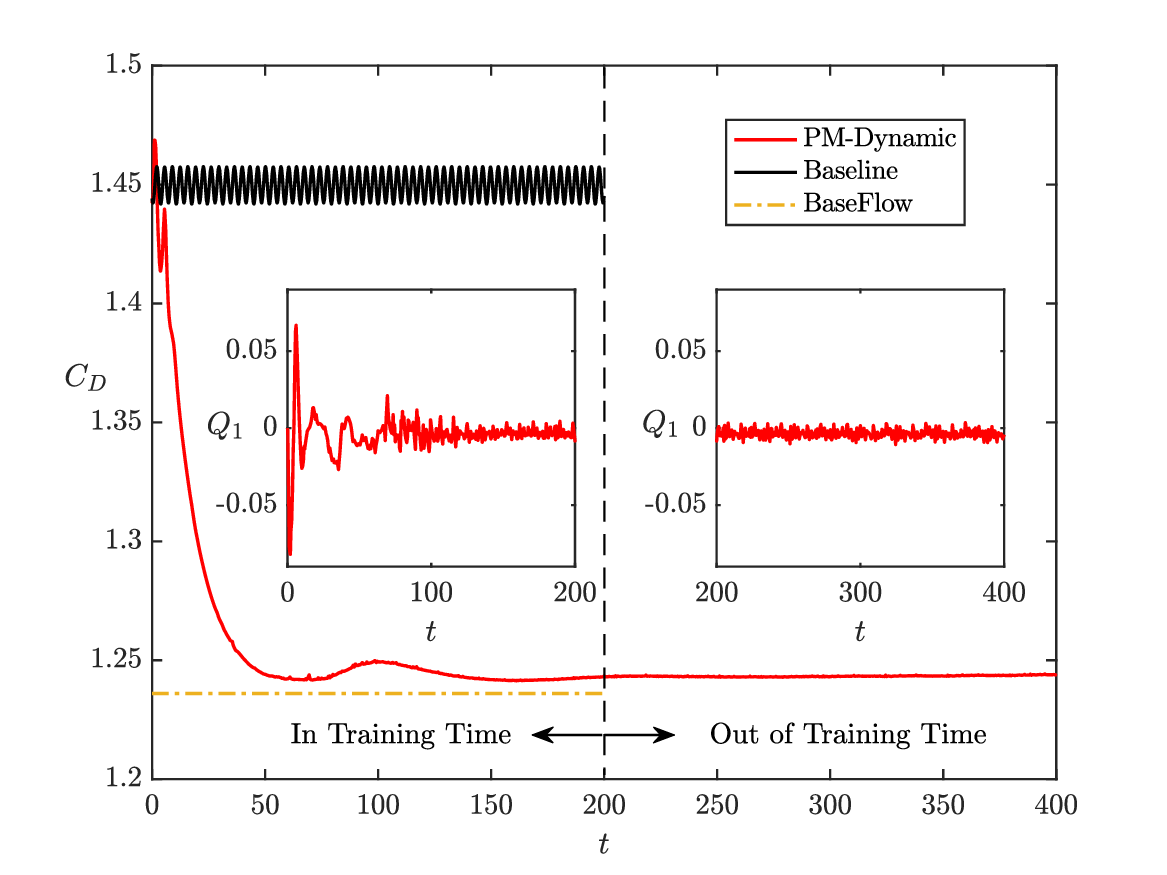}}
  \caption{A long evaluation for 400 non-dimensional time units of the RL-trained dynamic controller in a PM environment. Control starts at $t=0$. Solid curves show the controlled $C_D$ using TQC and ``Baseline'' without control. The dash-dotted curve corresponds to the baseflow $C_D$. The mass flow rate $Q_1$ is presented for $t\in [0,200]$ and $t\in [200,400]$ respectively.}
\label{fig:LongEva}
\end{figure}

In figure \ref{fig:LongEva}, the trained policy is tested for a longer time (400 time units) than training (200 time units) to show the control stability outside the training timeframe for the dynamic controller in the PM environment.
The initial condition of this long-run test is different compared to figure \ref{fig:TQC_FMPM}, indicating the adaptability of the controller to different initial conditions. Other parameters in this run are consistent with the results in figure \ref{fig:TQC_FMPM}.

The control performance and behaviour in this test are consistent with the results shown in figure \ref{fig:TQC_FMPM} both in the transient stage and asymptotic stage. The drag coefficient $C_D$ starts from the condition of steady vortex shedding and drops to the value of the stabilised flow in around 120 time units, with minor fluctuations. After training time (200 time units), the controller is still able to prevent triggering vortex shedding and preserve the drag coefficient near the baseflow values (minimum drag without vortex shedding). The behaviour of the controller is further presented in the subfigures of $Q_1$. The controller creates negligible random mass flow after stabilizing the vortex shedding due to the maximum entropy in training.

\section{Base flow simulation} \label{App:BaseFlow}

\begin{figure}
  \centerline{\includegraphics[width=14cm]{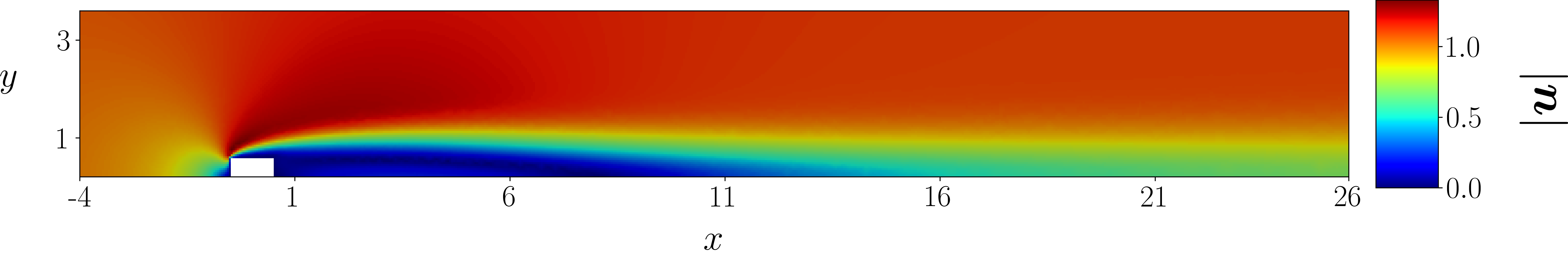}}
  \caption{Baseflow  (steady flow without vortex shedding) obtained with a half domain simulation, i.e. $y \in [0,12.5]$. A sub-domain of $y \in [0,3.5], x \in [-4,26]$ is plotted for demonstration. The symmetric boundary condition is applied on the $y = 0$ boundary. The mesh of the simulation is consistent with figure \ref{fig:Mesh}.}
\label{fig:Baseflow}
\end{figure}

The base flow corresponds to a steady equilibrium of the governing Navier-Stokes equations. This fixed point is unstable to infinitesimal perturbations, giving rise to vortex shedding. The base flow is obtained by simulating only half of the domain, as shown in figure \ref{fig:Baseflow}, which prevents antisymmetric vortex shedding. The boundary conditions are consistent with Eq. \req{eq:BCs} while a symmetric boundary condition is applied on the bottom boundary (symmetry line) of the domain, i.e. on $y = 0$. The symmetric boundary condition is imposed as $v = 0$, $\frac{\partial u}{\partial y} = 0$ and $\frac{\partial p}{\partial y} = 0$.

In this case, the vortex shedding is not triggered, as shown in the contour of figure \ref{fig:Baseflow}, and the only cause of the pressure drag is flow separation. Therefore, comparing the pressure drag in a full-domain simulation of uncontrolled flow, where the vortex shedding is triggered, with this base flow, the amount of pressure drag due to flow unsteadiness can be estimated. As only the unsteady component of pressure drag can be effectively reduced by flow control \citep{bergmann_optimal_2005}, the control performance can be evaluated with respect to this base flow (Eq.\req{eq:drag_reduction}). The drag coefficient of the half square body measures $C_{Dh} = 0.618$, and the base flow drag coefficient of the whole body can be obtained as $C_{Db} = 2C_{Dh} = 1.236$. 

